\documentclass[useAMS,usenatbib]{mn2e}
\usepackage{amsmath}
\usepackage{amssymb}
\usepackage{graphicx}
\usepackage{lscape}
\usepackage{listings}
\usepackage{hyperref}
\usepackage{amstext} 
\usepackage{array}   
\newcolumntype{L}{>{$}l<{$}} 

\newcommand{\likL}{\mathcal{L}}

\newcommand{\meth}{\mbox{CH$_{4}$ }}
 
\title[Retrievals of L dwarfs]{Retrieval of atmospheric properties of cloudy L dwarfs}
\author[Burningham et al]{Ben Burningham$^{1,2}$\thanks{E-mail:
    B.Burningham@herts.ac.uk}, M. S. Marley$^{1}$, M. R. Line$^{3,1}$, R. Lupu$^{1,4}$, C. Visscher$^{5,6}$, 
    \newauthor  C. V. Morley$^{7}$, D. Saumon$^{8}$,  R. Freedman$^{1,9}$\\
$^{1}$ NASA Ames Research Center, Mail Stop 245-3, Moffett Field, CA 94035, USA \\
$^{2}$ Centre for Astrophysics Research, School of Physics, Astronomy and Mathematics, University of Hertfordshire, Hatfield AL10 9AB \\
$^{3}$ School of Earth \& Space Exploration, Arizona State University, Tempe AZ 85287, USA \\
$^{4}$ Bay Area Environmental Research Institute, 625 2nd Street, Suite 209, Petaluma, CA 94952, USA\\
$^{5}$ Department of Chemistry, Dordt College, Sioux Center, IA 51250, USA \\
$^{6}$ Space Science Institute, Boulder, CO, USA \\
$^{7}$ Department of Astronomy, Harvard University, Cambridge, MA 02138, USA \\
$^{8}$ Los Alamos National Laboratory, P.O. Box 1663, MS F663, Los Alamos, NM 87545, USA \\
$^{9}$ SETI Institute, Mountain View, CA 94043, USA
}

\begin{document}
%
%
%
%


\def\aj{\rm{AJ}}                   
\def\araa{\rm{ARA\&A}}             
\def\apj{\rm{ApJ}}                 
\def\apjl{\rm{ApJ}}                
\def\apjs{\rm{ApJS}}               
\def\ao{\rm{Appl.~Opt.}}           
\def\apss{\rm{Ap\&SS}}             
\def\aap{\rm{A\&A}}                
\def\aapr{\rm{A\&A~Rev.}}          
\def\aaps{\rm{A\&AS}}              
\def\azh{\rm{AZh}}                 
\def\baas{\rm{BAAS}}               
\def\jrasc{\rm{JRASC}}             
\def\memras{\rm{MmRAS}}            
\def\mnras{\rm{MNRAS}}             
\def\pra{\rm{Phys.~Rev.~A}}        
\def\prb{\rm{Phys.~Rev.~B}}        
\def\prc{\rm{Phys.~Rev.~C}}        
\def\prd{\rm{Phys.~Rev.~D}}        
\def\pre{\rm{Phys.~Rev.~E}}        
\def\prl{\rm{Phys.~Rev.~Lett.}}    
\def\pasp{\rm{PASP}}               
\def\pasj{\rm{PASJ}}               
\def\qjras{\rm{QJRAS}}             
\def\skytel{\rm{S\&T}}             
\def\solphys{\rm{Sol.~Phys.}}      
\def\sovast{\rm{Soviet~Ast.}}      
\def\ssr{\rm{Space~Sci.~Rev.}}     
\def\zap{\rm{ZAp}}                 
\def\nat{\rm{Nature}}              
\def\iaucirc{\rm{IAU~Circ.}}       
\def\aplett{\rm{Astrophys.~Lett.}} 
\def\apspr{\rm{Astrophys.~Space~Phys.~Res.}}
\def\bain{\rm{Bull.~Astron.~Inst.~Netherlands}} 
\def\fcp{\rm{Fund.~Cosmic~Phys.}}  
\def\gca{\rm{Geochim.~Cosmochim.~Acta}}   
\def\grl{\rm{Geophys.~Res.~Lett.}} 
\def\jcp{\rm{J.~Chem.~Phys.}}      
\def\jgr{\rm{J.~Geophys.~Res.}}    
\def\jqsrt{\rm{J.~Quant.~Spec.~Radiat.~Transf.}}
\def\memsai{\rm{Mem.~Soc.~Astron.~Italiana}}
\def\nphysa{\rm{Nucl.~Phys.~A}}   
\def\physrep{\rm{Phys.~Rep.}}   
\def\physscr{\rm{Phys.~Scr}}   
\def\planss{\rm{Planet.~Space~Sci.}}   
\def\procspie{\rm{Proc.~SPIE}}   

\let\astap=\aap
\let\apjlett=\apjl
\let\apjsupp=\apjs
\let\applopt=\ao

\maketitle

\begin{abstract}
We present the first results from applying the spectral inversion technique in the cloudy L dwarf regime. Our new framework provides a flexible approach to modelling cloud opacity which can be built incrementally as the data requires, and improves upon previous retrieval experiments in the brown dwarf regime by allowing for scattering in two stream radiative transfer.  Our first application of the tool to two mid-L dwarfs is able to reproduce their near-infrared spectra far more closely than grid models. Our retrieved thermal, chemical, and cloud profiles allow us to estimate $T_{\rm eff} = 1796^{+23}_{-25}$~K and $\log g  = 5.21^{+0.05}_{-0.08}$ for 2MASS~J05002100+0330501 and for 2MASSW~J2224438-015852 we find $T_{\rm eff} = 1723^{+18}_{-19}$~K and $\log g = 5.31^{+0.04}_{-0.08}$, in close agreement with previous empirical estimates.
Our best model for both objects includes an optically thick cloud deck which passes $\tau_{cloud} \geq 1$ (looking down) at a pressure of around 5 bar. The temperature at this pressure is too high for silicate species to condense, and we argue that corundum and/or iron clouds are responsible for this cloud opacity. Our retrieved profiles are cooler at depth, and warmer at altitude than the forward grid models that we compare, and we argue that some form of heating mechanism may be at work in the upper atmospheres of these L dwarfs. We also identify anomalously high CO abundance in both targets, which does not correlate with the warmth of our upper atmospheres or our choice of cloud model, and find similarly anomalous alkali abundance for one of our targets. These anomalies may reflect unrecognised shortcomings in our retrieval model, or inaccuracies in our gas phase opacities.
 \end{abstract}


\section{Introduction}
\label{sec:intro}
 
The study of brown dwarf atmospheres has been dominated by fitting the spectral predictions of grids of self-consistent 1-dimensional radiative-convective equilibrium atmosphere models (hereafter ``grid models") to optical and near-infrared spectroscopic and photometric observations,  and thence inferring properties of the target under study \citep[e.g.][]{cushing2008, manjavacas2014,rice2015}. In a handful of cases, so-called ``benchmark systems", properties for substellar targets in binary systems have been estimated by association with a stellar primary, and the quality of fit used to critically evaluate the grid models \citep[e.g.][]{ben2009,ben2011b,ben2013}.  These approaches are hindered by the difficulty of identifying the factors that drive poor matches between observation and synthetic spectra from a range of possibilities that arise from the inherent complexity of the grid models and atmospheres to which they relate. 

The lineage of these forward model grids can typically be traced back to the pioneering efforts of a handful of teams who either adapted Solar system planetary atmosphere codes to higher temperature and gravity regimes \citep[e.g. ][]{marley1996,saumon2012}, or extended cool stellar models downwards \citep[e.g.][]{allard1995,btsettlCS16,burrows1993,burrows2006,tsuji1996,tsuji2002}.  Both approaches rely on an iterative process of refining the thermal structure under the assumption of radiative-convective equilibrium.  An initial guess for the T(P) profile is made, opacities calculated and convective and radiative energy transport determined to estimate a new profile. This continues until the profile converges.This process typically employs a combination of elements, including: a thermochemical model to self-consistently estimate gas volume mixing ratios (and thus opacity) as a function of (T, P) on the profile; a cloud model to determine condensate opacity along the profile; a convection model; a radiative transfer solver.  Each of these elements involves a variety of decisions regarding appropriate approximations, as does the iterative process itself.  A comprehensive review of the current state-of-the-art of grid modelling may be found in \citet{marley2015}, which also highlights some key areas of uncertainty in these approaches.

Various shortcomings in the predictions from grid models have been identified both from detailed studies of small numbers of benchmark systems, and also through wider comparisons of the broadband photometric colours of the population across the L~and T~spectral sequences. For example, the BT-Settl model grids \citep{btsettlCS16} have difficulty reproducing the general slope of near-infrared spectra of $L$~dwarfs and specifically the shape of the $H$~band peak \citep{manjavacas2014,manjavacas2016}, although they are able to reproduce late-T dwarf spectral energy distributions quite well \citep[e.g.][]{ben2011b}. In the $L$ dwarf regime, there are a number of objects that show discrepancies of several 100K between the $T_{\rm eff}$ inferred through grid model fitting, and that found from bolometric luminosity and radius estimates \citep[including an object we consider in this paper: 2MASSW~J2224438-015852; ][]{stephens2009}. This presumably arises from either the grid models' poor fit to the L~dwarfs' SED due to some unforeseen property of the targets, or some unaccounted for physical process(es).  

The population of L~dwarfs in the solar neighbourhood displays a considerable diversity of spectrophotometric properties which have been variously associated with low-gravity \citep[red $J-K$ colours; e.g. ][]{faherty2013,allers2013}, low-metallicity and/or high-gravity \citep[blue $J-K$ colours; e.g. ][]{faherty2009}, unusual cloud properties \citep[blue $J-K$ colours; e.g. ][]{burgasser2008,cushing2010,marley2010}. Given the relatively narrow range of compositions (e.g. $\pm 0.5$ dex) and gravities ($\pm 1$ dex) that solar neighbourhood brown dwarfs are expected to occupy, this diversity likely reflects the complexity inherent to the interplay of chemistry, atmospheric dynamics and clouds.  This complexity is akin to that expected for exoplanets, and already seen in Solar system atmospheres. Cloudy L~dwarfs are thus outstanding laboratories for examining and testing techniques to explore this complexity and relate observational groups and behaviours to particular drivers of atmospheric chemistry and physics.

Spectral inversion techniques (or atmospheric retrievals) provide the opportunity to confront forward grid models with observations in a more nuanced manner than simply comparing model spectra to data. Originally developed for remote sensing Earth's \citep[e.g.][]{rodgers1976,rodgers2000} and Solar system planetary atmospheres \citep[e.g.][]{irwin2008}, retrievals have been used to study transiting exoplanets \citep[e.g.][]{barstow2013,line2013}, directly imaged planets \cite{lee2013}, and simple cloud-free brown dwarf atmospheres \citep{line2014,line2015}.  In the case of cloudy brown dwarfs and directly imaged exoplanets, this technique allows for empirical constraints to be placed directly on the gas volume mixing ratios, thermal profile,  and cloud location and opacity. These can be compared directly to the outputs of grid models, allowing detailed examination and evaluation of differing approaches to approximating the atmospheric physics of brown dwarfs and giant exoplanets.

In this paper we introduce a new framework for performing retrievals in the cloudy L~dwarf regime, and present its first application to two mid-L~dwarfs. Performing retrievals in this regime is more challenging than in the cloud-free late-T regime that has been explored by \citet{line2014,line2015} with this technique. Whereas T~dwarf spectra have bright peaks in the opacity windows between the deep absorption bands due to water and methane, mid-L~dwarf spectra have weaker contrast in these regions, which is symptomatic of a photosphere that spans a relatively narrower temperature and pressure range. As a result careful validation is required to ensure that our retrieved cloud and thermal profiles stand up to scrutiny.  

In Section~\ref{sec:brew} we introduce our framework and outline its key details. In Section~\ref{sec:fakedata} we describe our validation of the technique using simulated targets. Section~\ref{sec:2m2224} describes the first application of this tool to two mid-L~dwarfs, and Section~\ref{sec:conc} rounds off our discussion of our results and summarises our conclusions. 

\section{A tool for retrieving properties of cloudy atmospheres}
\label{sec:brew}

The two principal structures in a retrieval framework are the forward model and the retrieval model. The forward model produces the spectrum based on a set of atmospheric parameters, which are typically a mixture of parameters to be retrieved and others for which values are assumed or otherwise pre-determined. The retrieval model performs the process of evaluating the goodness of fit for forward model outputs and proposing new values to effectively sample the posterior probability distribution of the parameters to be retrieved. In the following sections we outline the key features of each of these elements.

\subsection{The forward model}
\label{subsec:forward}

The essence of the forward model can be reduced to a set of three key elements: the radiative transfer solver, the thermal profile, and the opacity and scattering properties in each layer as a function of wavelength. 
Our forward model solves the radiative transfer to evaluate the emergent flux from a layered atmosphere in the two stream source function technique of \citet{toon1989}, including scattering, as first introduced by \citet{mckay1989} and subsequently used by e.g. \citet{marley1996,sm08,morley2012}. We set up a 64 layer atmosphere (65 levels) with geometric mean pressures in the range $\log P(bar) = -4$ to 2.3, spaced at 0.1~dex intervals.  

\subsubsection{Thermal profile}

The thermal profile can be parameterised in a number of ways, or not parameterised at all and the temperature of each layer retrieved independently. The latter option must be treated with caution since the data alone are unlikely to adequately constrain the profile shape, and unphysical solutions with significant discontinuity are likely to result without applying some sort of smoothing \citep[e.g.][]{irwin2008, line2014}. 
An alternative is to sample the atmospheric profile at wider pressure steps than are required for accurate radiative transfer calculations, and then interpolate the full profile from these. This removes the risk of small scale fluctuations in the profile, and reduces the number of parameters for the retrieval. However, even low resolution interpolated profiles are prone to oscillatory morphologies, and smoothing is thus still desirable. In an attempt to avoid this requirement, \citet{line2015} used a low-resolution thermal profile and spline-interpolation to the full resolution pressure scale, but penalised the second derivative of the final curve in the retrieval. This has the effect of minimising the jaggedness of the profile unless the benefit to improving the fit is significant. This method has been tested within our framework, however early experiments in the L~dwarf regime found that the data rarely justified anything other than an entirely linear profile, and the retrievals did not converge. This is due to the spectral contrast in the data being unable to overcome the penalisation of the second derivative in the Gaussian Process used by \citet{line2015} to drive the profile away from the linear form that it is inherently biased towards.

Instead, we have adopted a parameterisation for our thermal profile, further reducing the number of parameters required to describe the thermal structure. The five and six variable parameterisations put forward by \citet{madhu2009} are able to fit most physically plausible thermal profiles. However, since they consist of a set of joined exponential curves this method will not retrieve profiles with local discontinuities. In the \citet{madhu2009} scheme, the atmosphere is split into three zones, whose pressure and temperature are related as:

\begin{equation}
\begin{aligned}
P_{0} < P < P_{1}: P  = P_{0} e^{\alpha_{1}(T - T_{0})^{\frac{1}{2}}}   \hfill (\text{Zone 1})\\
P_{1} < P < P_{3}: P  = P_{2} e^{\alpha_{2}(T - T_{2})^{\frac{1}{2}}}   \hfill (\text{Zone 2})\\
P > P_{3} : T = T_{3}  \hfill (\text{Zone 3})
\end{aligned}
\label{eqn:madhu}
\end{equation}
where $P_{0}$, $T_{0}$ the pressure and temperature at the top of the atmosphere, which becomes isothermal with temperature $T_{3}$ at pressure $P_{3}$.
In its most general form, a thermal inversion occurs when $P_{2} > P_{1}$. Since $P_{0}$ is fixed by our atmospheric model, and continuity at the zonal boundaries allows us to fix two parameters, we consider six free parameters $\alpha_{1}$, $\alpha_{2}$, $P_1$, $P_2$, $P_3$, and $T_3$.  If we rule out a thermal inversion by setting $P_{2} = P_{1}$ \citep[see Figure 1, ][]{madhu2009},  we can further simplify this to five parameters $\alpha_{1}$, $\alpha_{2}$, $P_1$, $P_3$, $T_3$.  

We note that although the deepest zone of this parameterisation is isothermal, this does not force our profile to be isothermal. The ($P_{3}, T_{3}$) point essentially acts as an anchor for our profile, and its location helps set the gradient of the profile at depth. There is no requirement for $P_{3}$ to lie within our model atmosphere for radiative transfer purposes, and it may be located at much greater depth than can be ``seen''.  An alternative approach would be to set some functional form for the $P > P_{3}$ region.  However, all options where this could provide a meaningful difference from the isothermal choice in a retrieval context would require additional parameters, and any parameter-free options would be essentially equivalent to our current choice. 

We have verified that this choice of parameterisation does not bias against the types of profiles expected for brown dwarf atmospheres by ensuring that we are able to fit a range of cloudy and cloud-free grid model thermal profiles from \citet{sm08}. Indeed, we use these parameterised fits to initialise our MCMC chains in Section~\ref{sec:2m2224}. In Section~\ref{sec:fakedata} we establish this further by successfully retrieving thermal profiles for simulated data generated using the \citet{sm08} grid profiles.

\subsubsection{Gas opacity}

For simplicity, we assume uniform-with-altitude mixing ratios for absorbing gases in all the cases considered here, although our code is designed to accept layer-by-layer varying gas mixing ratios if desired. 
We calculate layer optical depths due to absorbing gases using high-resolution cross-sections sampled at 1cm$^{-1}$ \citep[as justified by ][ for Spex data]{line2016} taken from the compendium of \citet{freedman2008,freedman2014}.

Here we use a new calculation of atomic line absorption due to neutral sodium and potassium using the VALD3 line list \citep[][\footnote{\url{http://vald.astro.univie.ac.at/~vald3/php/vald.php}}]{ryabchikova2015}. The line profiles are generally assumed to be Lorentzian, with no line cutoff in strength or frequency applied. The line width is calculated from the Van der Waals broadening theory for collisions with $H_{2}$ molecules using the coefficient tabulated in the VALD3 data base when available or from the full theory otherwise.

The D resonance doublets of NaI ($\sim 0.59~\mu$m) and KI ($\sim 0.77~\mu$m) can become extremely strong in the spectra of brown dwarfs, with line profiles detecable as far as $\sim$3000~cm$^{-1}$ from the line centre in T dwarfs \citep[e.g. ][]{burrows2000,liebert2000,marley2002,king2010}. Under these circumstances, a Lorentzian line profile becomes woefully inadequate in the line wings and a detailed calculation is required. For these two doublets, we have implemented line wing profiles based on the unified line shape theory \citep{nallard2007a,nallard2007b}. The tabulated profiles (Allard N., private communication) are calculated for the D1 and D2 lines of Na I and K I broadened by collisions with H$_{2}$ and He, for temperatures in the $500 - 3000$~K range and perturber (H$_{2}$ or He) densities up to $10^{20}$ cm$^{-3}$. Two collisional geometries are considered for broadening by $H_{2}$. The profile within 20~cm$^{-1}$ of the line centre is Lorentzian with a width calculated from the same theory. 

We have computed a new water opacity database using the UCL (ExoMol) linelists, including the main isotope from 2006 and newer data for H$_{2}^{17}$O and H$_{2}^{18}$O. We normally do not include HDO in our standard water opacities, but can be calculated separately. The partition functions have been updated to the UCL values, and the line widths combine UCL data for He and H$_2$ broadening. We combine the two widths, each with its own temperature dependence, using an assumed proportion of 0.85H$_{2}$ + 0.15He to give the actual effective broadening width for each line.  We also performed the investigations described in this paper using the \citet{partridge1997} H$_{2}$O opacities and found no significant differences in our results.

The line opacity cross-sections are tabulated across our temperature-pressure regime in 0.5~dex steps for pressure, and with temperature steps ranging from 20~K to 500~K as we move from 75~K to 4000~K. We then linearly interpolate to our working pressure grid.  We also include continuum opacities for H$_{2}$-H$_{2}$ and H$_{2}$-He collisionally induced absorption, using the updated cross-sections from  \citet{richard2012} and \citet{saumon2012}.  We also include Rayleigh scattering due to H$_{2}$, He and CH$_{4}$ but neglect the remaining gases.

\subsubsection{Clouds}

We have designed our forward model with a variety of options for cloud treatments with a range of complexities, which we describe below. We discuss the clouds' optical depths only terms of extinction. All cloud options include a free-to-vary single-scattering albedo ($w_{0}$), but we assume isotropic scattering and fix the value of the asymmetry parameter to zero.  

The simplest proscription we consider is a 3-parameter model where the cloud is assumed to be grey and very optically thick with only the top of the cloud deck visible. We define the cloud deck as the pressure at which the cloud optical depth passes unity (looking down). The cloud opacity to drops off in $d\tau/dP$ towards lower pressures than the cloud deck and increases to higher pressures, with $d\tau/dP \propto \exp(\Delta P / \Phi)$ where $\Delta P$ is the height above or below the optically thick level of the cloud deck, and  $\Phi$ is the decay scale of the cloud in bars. This is parameterised in the retrieval by a decay scale expressed in $d\log P$.  The cloud is thus characterised with three parameters: the pressure level where the optical depth passes unity, the decay scale pressure, and the single-scattering albedo. 

The next level of complexity treats the cloud as a slab parameterised by total optical depth of the cloud, cloud top pressure, and cloud thickness (in $d\log P$). For the slab cloud we force $d\tau / dP \propto P$. With an extra parameter, both of these cases can be adapted to a non-grey cloud with an optical depth that follows a power-law, $\tau = \tau_{0} \lambda^{\alpha}$, where $\tau_{0}$ is the cloud optical depth at 1$\micron$.

\subsection{The retrieval model}
\label{subsec:retrieval}

\begin{figure*}
\hspace{-0.8cm}
\includegraphics[width=550pt]{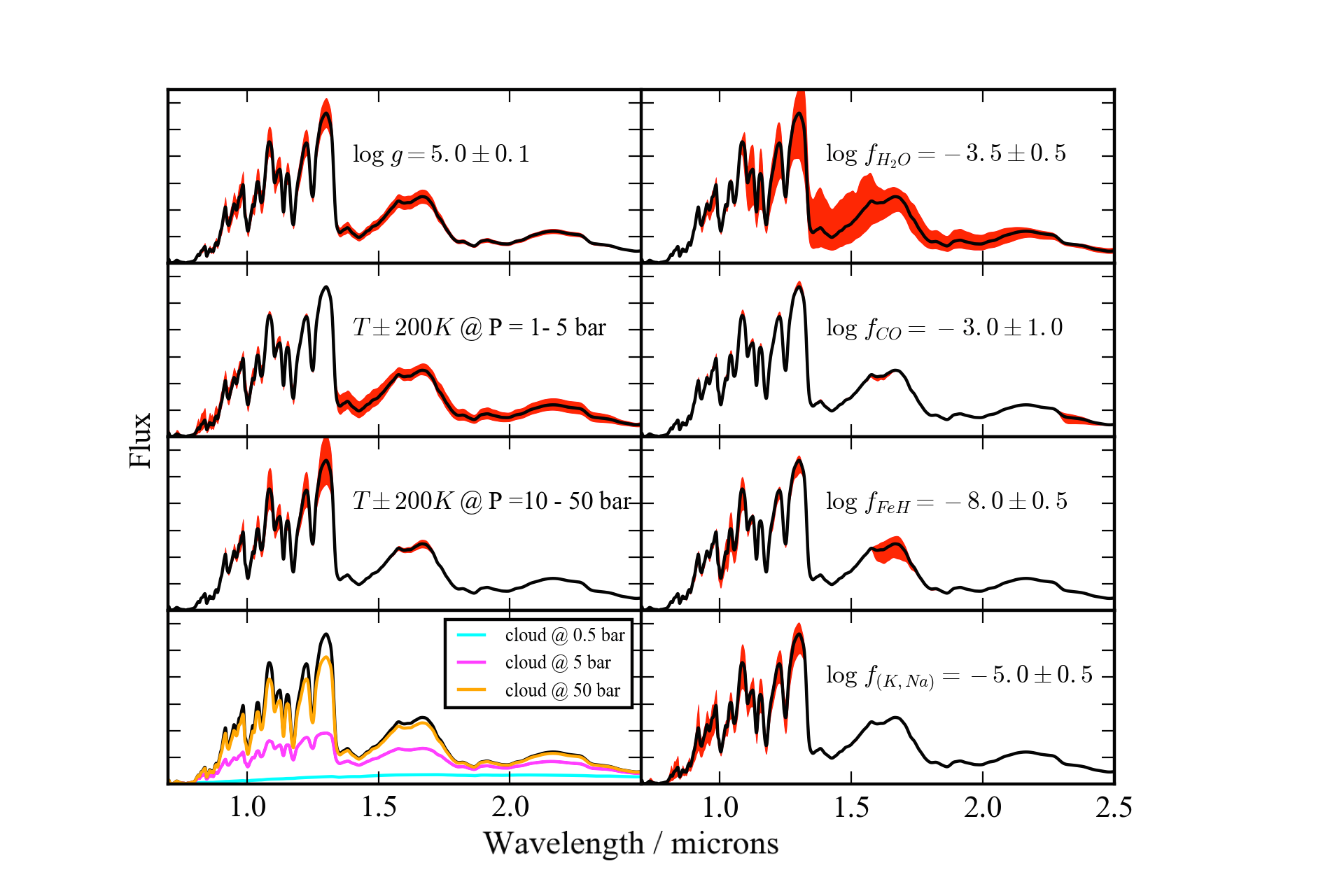}
\caption{A demonstration of the impact of varying different elements of the forward model state-vector on the emergent spectrum. The reference case here is a cloud-free atmosphere with a thermal profiles taken from a $T_{\rm eff} = 1700$K, $\log g = 5.0$ \citet{saumon2012} forward-grid model. The impact on the resultant spectrum of varying the gravity and the temperature at different depths, adding a cloud and varying the gas abundances are indicated on each panel.  The cloud that we add in this example is the simple grey cloud deck scenario where the depth specifies the pressure at which the $\tau_{cloud} > 1$, and we set a decay height of 1 dex.
\label{fig:influence}}
\end{figure*}

The retrieval process depends on the differing effects that changes in the elements of the parameter set that is passed to the forward model have on the resultant spectrum. Figure~\ref{fig:influence} illustrates the impact of adjusting some of these parameters on the resultant spectrum. The process of optimising the forward model's fit to the data by varying the input parameter set, or state-vector, takes place within a Bayesian framework.

Bayes' theorem, applied to parameter estimation, provides a route to calculating the probability of a set of parameters' ({\bf x}) truth given some some data ({\bf y}), $p({\bf x} | {\bf y})$ (the ``posterior probability''), as:\\

\begin{equation}
p({\bf x} | {\bf y}) = \frac{\likL({\bf y} | {\bf x}) p({\bf x})}{p({\bf y})}
\label{eqn:bayes}
\end{equation}
where $\likL({\bf y} | {\bf x})$ is the likelihood that quantifies how well the data match the model, $p({\bf x})$ is the prior probability on the parameter set, and $p({\bf y})$ is the probability of the data marginalised over all parameter values, known as the marginal likelihood or Bayesian evidence.  In the case of parameter estimation the Bayesian evidence simply acts as a normalisation factor. So, to map out the distribution of $p({\bf x} | {\bf y})$ and find the best set of parameters given the data, we need only consider the two terms on top of Equation~\ref{eqn:bayes}.

To sample the posterior distribution we use the \textsc{emcee} affine-invariant implementation of the Markov chain Monte Carlo method \citep{emcee}.  The {\sc emcee} samples the log of the posterior, and we use a log-likelihood function to assess the fit of the data to the model, as suggested by \citet{emcee}:

\begin{equation}
\ln \likL({\bf y} | {\bf x}) = - \frac{1}{2} \sum_{i = 1}^{n} \frac{(y_{i} - F_{i}(\bf x))^2}{s^{2}_{i}} - \frac{1}{2} \ln(2 \pi s^{2}_{i}) 
\label{eqn:lnlik}
\end{equation}
where the index $i$ refers to the $i$th of $n$ spectral flux points, $y_i$, with errors $s_i$,  which are compared to the forward model fluxes $F_i$ for the current parameter set ${\bf x}$. Following suggestions in the {\sc emcee} documentation, we inflate our errors using a tolerance parameter to allow for unaccounted for sources of uncertainty. Our data error, $s_i$, is given by: 

\begin{equation}
s_{i}^{2} = \sigma_{i}^{2} + 10^{b}
\end{equation}
where $\sigma_i$ is the measured error for the $i$th flux measurement, and $b$ is our tolerance parameter, which is retrieved \citep{emcee,hogg2010,tremaine2002}.

We have used uniform priors for all elements in the state-vector. The strongest priors we apply are on the radius and mass of the target under consideration. Though these two items are not retrieved directly, they combine to set a limit on the gravity.  The radius is determined from the scaling factor required to match the absolute flux level from the forward model, which is calculated for the top of the atmosphere (ToA flux), to the data, and the measured parallax for the target (scale factor = $R^{2}/D^{2}$).  We restrict the radius to within a very broad range of physically plausible values of 0.5 -- 2.0 $R_{Jup}$. The mass is restricted to between 1~$M_{Jup}$ and 80~$M_{Jup}$. These limits are set by reference to the \citet{sm08}, COND \citep{baraffe03} and DUSTY \citep{chabrier2000,baraffe2002} substellar evolutionary models. Both model grids predict $T_{\rm eff} > 2200$K for objects with 80$M_{Jup}$ at ages of 10~Gyr, and $T_{\rm eff} < 1000$K for objects with 1~$M_{Jup}$ and ages of 1~Myr. This range of ages comfortably encompasses the range of ages expected for members of the Galactic disk population at a broad range of $T_{\rm eff}$ bracketing a plausible range for mid-L~dwarfs considered here. 

In the cases examined here, our temperature profile is parameterised using 5 parameters (Eq. 1). We do not constrain these directly, but they are effectively restricted by a uniform prior on the resulting layer temperatures, $T_{i}$, which we restrict to between 0~K and 4000~K. Our full set of priors are given in Table~\ref{tab:priors}. 

\begin{table*}
\begin{tabular}{c  c}
\hline
Parameter & Prior \\
\hline
gas volume mixing ratio & uniform, $\log f_{gas} \geq -12.0$, $\sum_{gas}{f_{gas}} \leq 1.0$ \\
thermal profile:  $\alpha_{1}, \alpha{2}, P1, P3, T3$ & uniform, constrained by $0.0~{\rm K} < T_{i} < 5000.0~{\rm K}$ \\
scale factor, $R^{2} / D^{2}$ & uniform, constrained by $0.5 R_{Jup} \leq R \leq 2.0 R_{Jup}$ \\
gravity, $\log g$ & uniform, constrained by $1M_{Jup}  \leq gR^{2} / G \leq 80M_{Jup}$\\ 
cloud top$^{1}$ & uniform, $-4 \leq \log P_{CT}  \leq +2.3$ \\
cloud decay scale$^{2}$ & uniform, $0 < \log \Delta P_{decay} < 7$ \\
cloud thickness$^{3}$ & uniform, constrained by $\log P_{CT} \leq \log (P_{CT} + \Delta P) \leq 2.3$\\
cloud total optical depth (extinction)$^{3}$ & uniform, $\tau_{cloud} \geq 0.0$ \\
single scattering albedo, $w_{0}$ & uniform, $0. \geq w_{0} \leq 1.0$\\ 
wavelength shift & uniform, $-0.01< \Delta \lambda < 0.01 \micron$ \\
tolerance factor & uniform, $\log (0.01 \times min(\sigma_{i}^{2})) \leq b \leq \log(100 \times max(\sigma_{i}^{2}))$ \\
\hline
\end{tabular}
\caption{Notes: 1) for an optically thick cloud deck this is the pressure where $\tau_{cloud} = 1$, for a slab cloud this is the top level of the slab; 2) decay height for cloud deck above the $\tau_{cloud} = 1.0$ level; 3) thickness and $\tau_{cloud}$ only retrieved for slab cloud.   \label{tab:priors} }
\end{table*}

\section{Test cases}
\label{sec:fakedata}
As an initial test, we have validated our framework in the cloud-free regime by retrieving the atmospheric properties of the benchmark T~dwarf G570D. For the purposes of this validation we used the same alkali opacities as employed by \citet{line2015}, and our retrieved thermal profile, gas abundances and global properties were a close match to their results. The posterior distributions for the profile and gas abundances are given in the Appendix.

To validate our tool in the L~dwarf regime, we now present the results of retrievals of simple test cases based on spectra calculated using our forward model for a set of simple L~dwarf-like atmospheres.  These synthetic spectra were calculated with noise added assuming SNR=50 at the $J$ band flux peak, and convolved to a uniform resolution with FWHM = 0.005$\mu$m.  We have used the thermal profile from a $T_{\rm eff} = 1700$K, $\log~g = 5.0$ self-consistent grid model from \citet{sm08}, but have included only a simple set of 8 gases with arbitrarily selected mixing ratios, given in Table~\ref{tab:testBDgases}. 

\begin{table*}
\begin{tabular}{c L L L L L L L L}
\hline
Parameter & \log f_{H_{2}O }& \log f_{CO} & \log f_{TiO} & \log f_{VO} & \log f_{CrH} & \log f_{FeH} & \log f_{Na+K} & \log~g  \\
\hline
Input & -3.5 &  -3.0 & -7.5 & -8.5 & -8.0 & -8.0 & -5.0 & 5.0 \\
\hline
\end{tabular}
\caption{Input values for gas volume mixing ratios and gravity used for our test cases. \label{tab:testBDgases} }
\end{table*}

\subsection{Case 1}
 In Case 1, we have added a simple grey cloud of optical depth unity, with its cloud top at 1 bar, and a thickness of 0.1 dex.  For this case we explore our ability to retrieve the correct atmospheric parameters in both grey cloud and cloud-free scenarios. We initialised our emcee-MCMC runs with 16 walkers per parameter (19 parameters for grey cloud and 15 parameters for cloud-free cases) with values drawn from tight gaussian balls centred on values offset from the true values by small arbitrary amounts. The only exceptions were the cloud optical depth, physical thickness and albedo, which were initialised as uniform distributions between zero and one.  After 10000 steps of burn-in the sampler was reset and run for 30000 steps. We draw our results from the last 2000 steps, comprising $32000 \times N_{dim}$ independent samples. 

\begin{figure}
\hspace{-0.8cm}\includegraphics[width=290pt]{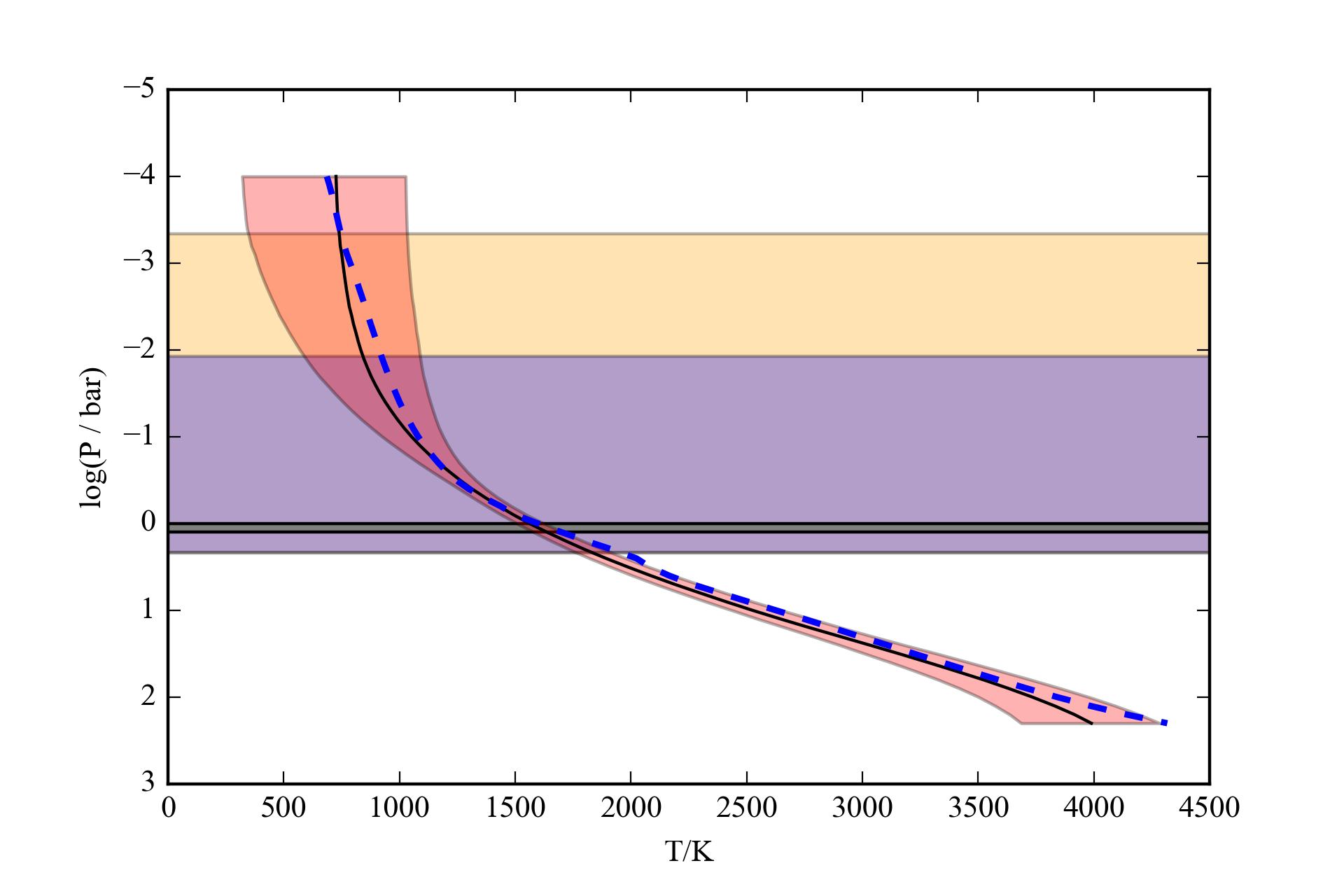}
\caption{Comparison of our retrieved thermal profile and cloud location with the input for our test case 1. Our median retrieved profile is indicated by a solid black line, and its 65\% confidence interval is denoted by red shading. The input profile is shown with a blue dashed line, and the input cloud location is shaded grey. The median retrieved cloud is shaded purple, and its 68\% confidence range is shaded yellow.
\label{fig:test1_cld}}
\end{figure}

\begin{figure}
\hspace{-0.8cm}\includegraphics[width=290pt]{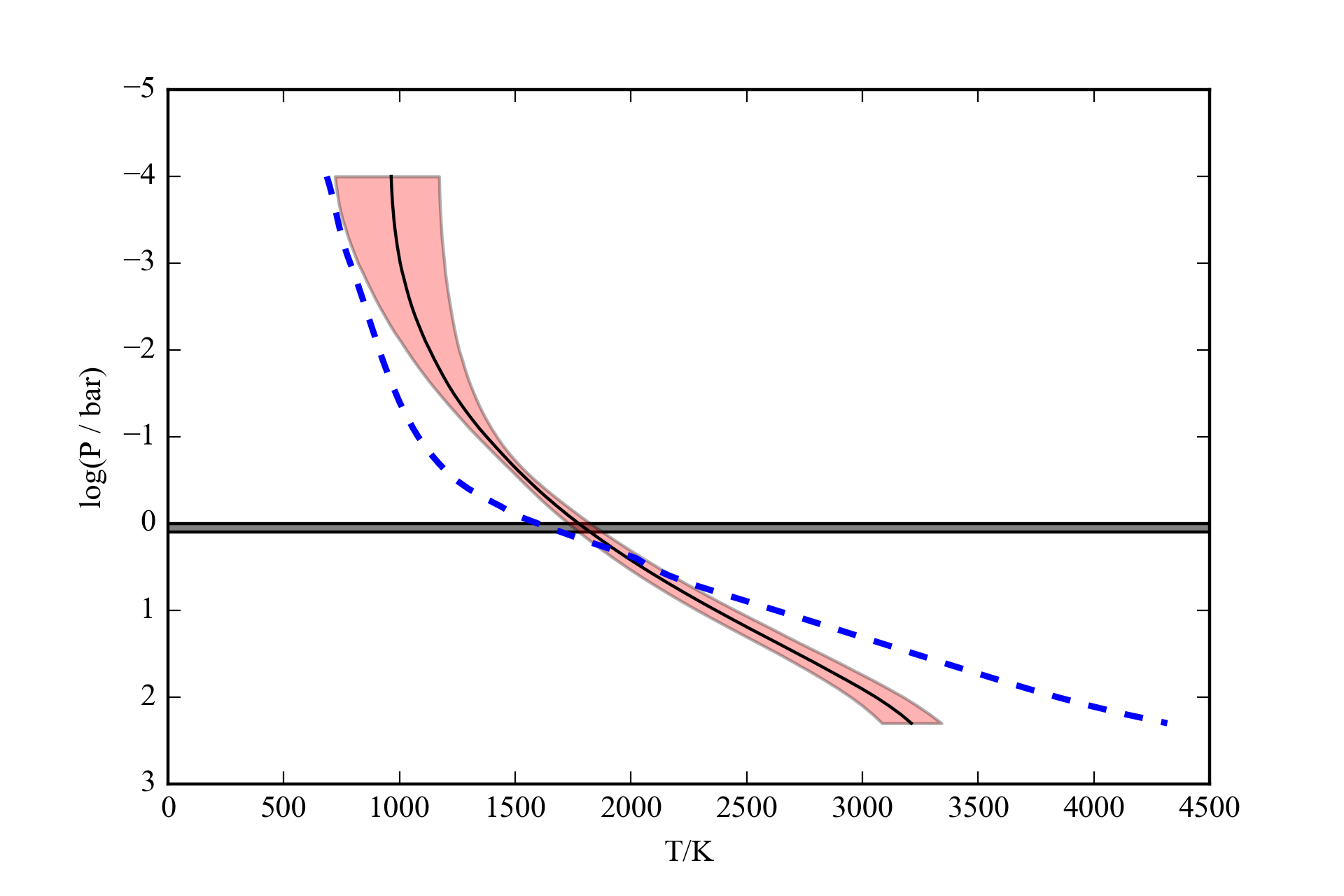}
\caption{Comparison of our retrieved thermal profile with the input for our test case 1, retrieved under a cloud-free assumption. Our median retrieved profile is indicated by a solid black line, and its 65\% confidence interval is denoted by red shading. The input profile is shown with a blue dashed line, and the input cloud location is shaded grey. 
\label{fig:test1_nc}}
\end{figure}

\begin{figure*}
\hspace{-0.8cm}\includegraphics[width=500pt]{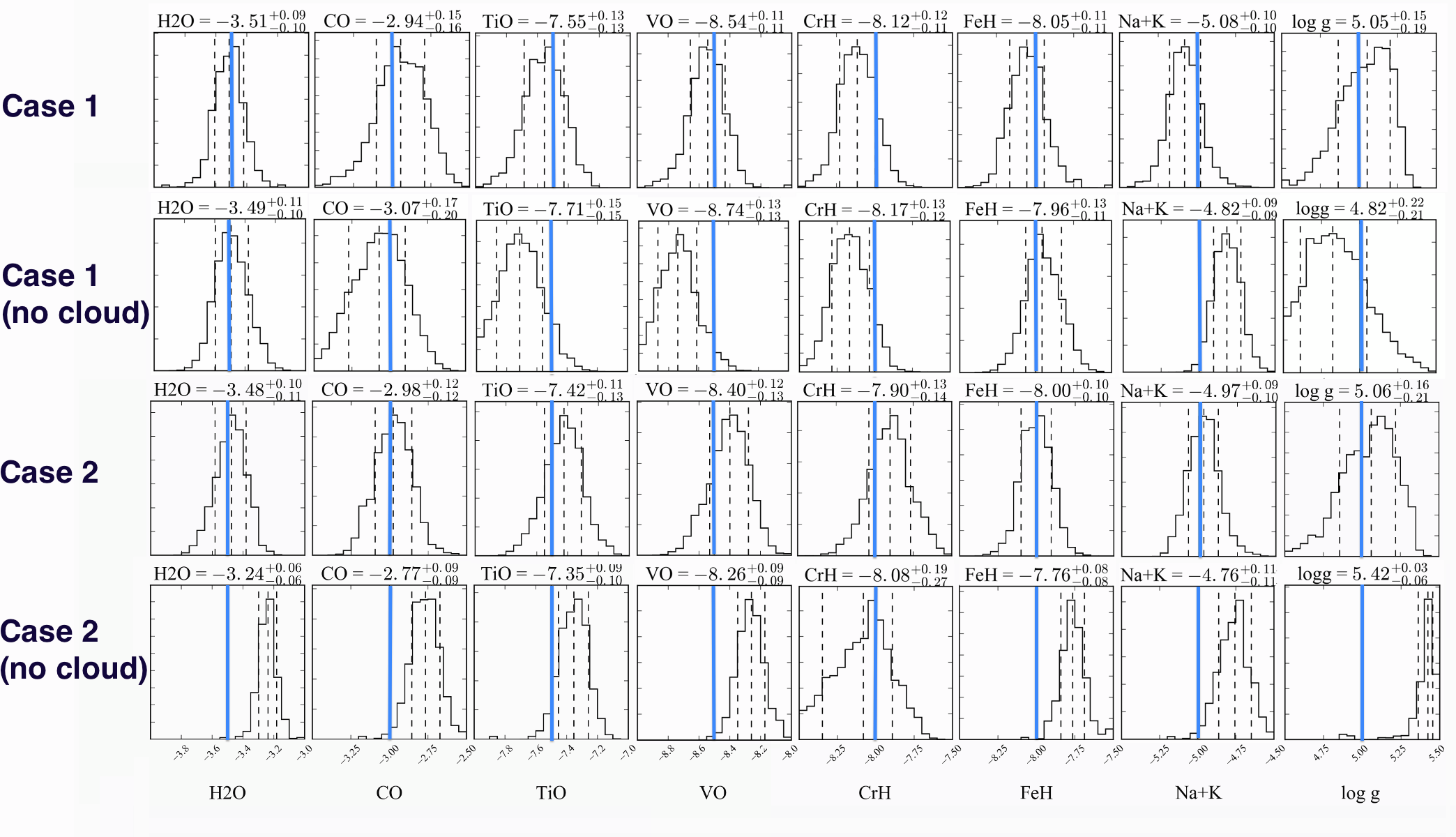}
\caption{Marginalised posterior distributions for the retrieved gas abundances in our two test cases. The input values are shown as blue lines, the median retrieved values and 1$\sigma$ intervals are shown as dashed lines on each histogram.
\label{fig:gasstack}}
\end{figure*}

Figures~\ref{fig:test1_cld} and ~\ref{fig:test1_nc} show our retrieved thermal profiles for case 1 under cloudy and cloud-free retrieval assumptions. Our retrieved profiles match the input profile closely under the cloudy assumption, and the retrieved gas fractions match the input values (Figure~\ref{fig:gasstack}). The retrieved cloud optical depth matches the input value well, $\tau_{cloud} = 1.01^{+0.24}_{-0.18}$. Our median retrieved cloud location is a good match for the input. Although the cloud physical thickness is not well constrained, the pressure level where $\tau_{cloud} = 1.0$ is well constrained and lies within 0.3~dex of the input cloud base.  


The retrieved profile for Case 1 under the cloud-free assumption (Figure~\ref{fig:test1_nc}) does not follow the input profile so closely, showing a more isothermal profile that is warmer at lower pressures than the approximate location of the cloud in the input model, and cooler at higher pressures. This is similar to the profile morphology suggested by \citet{tremblin2015,tremblin2016} for cloud-free L~dwarfs.  The retrieved gas abundances and gravity are similar to those retrieved under the cloudy assumption, and the input values, with the exception of the alkali abundance which is significantly overestimated in this case.  
To select a preferred retrieval model for this test case we apply the Bayesian Information Criterion (BIC). \citet{kass1995} provide the following intervals for selecting between two models under the BIC:

\begin{itemize}
  \item $0  < \Delta BIC < 2$: no preference worth mentioning;
  \item $2  < \Delta BIC < 6$: postive;
  \item $6  < \Delta BIC < 10$: strong;
 \item $10  < \Delta BIC$: very strong.
\end{itemize}
In these two cases, we find that our cloudy model is strongly preferred with a $\Delta BIC =11$.

\subsection{Case 2}

\begin{figure}
\hspace{-0.8cm}\includegraphics[width=290pt]{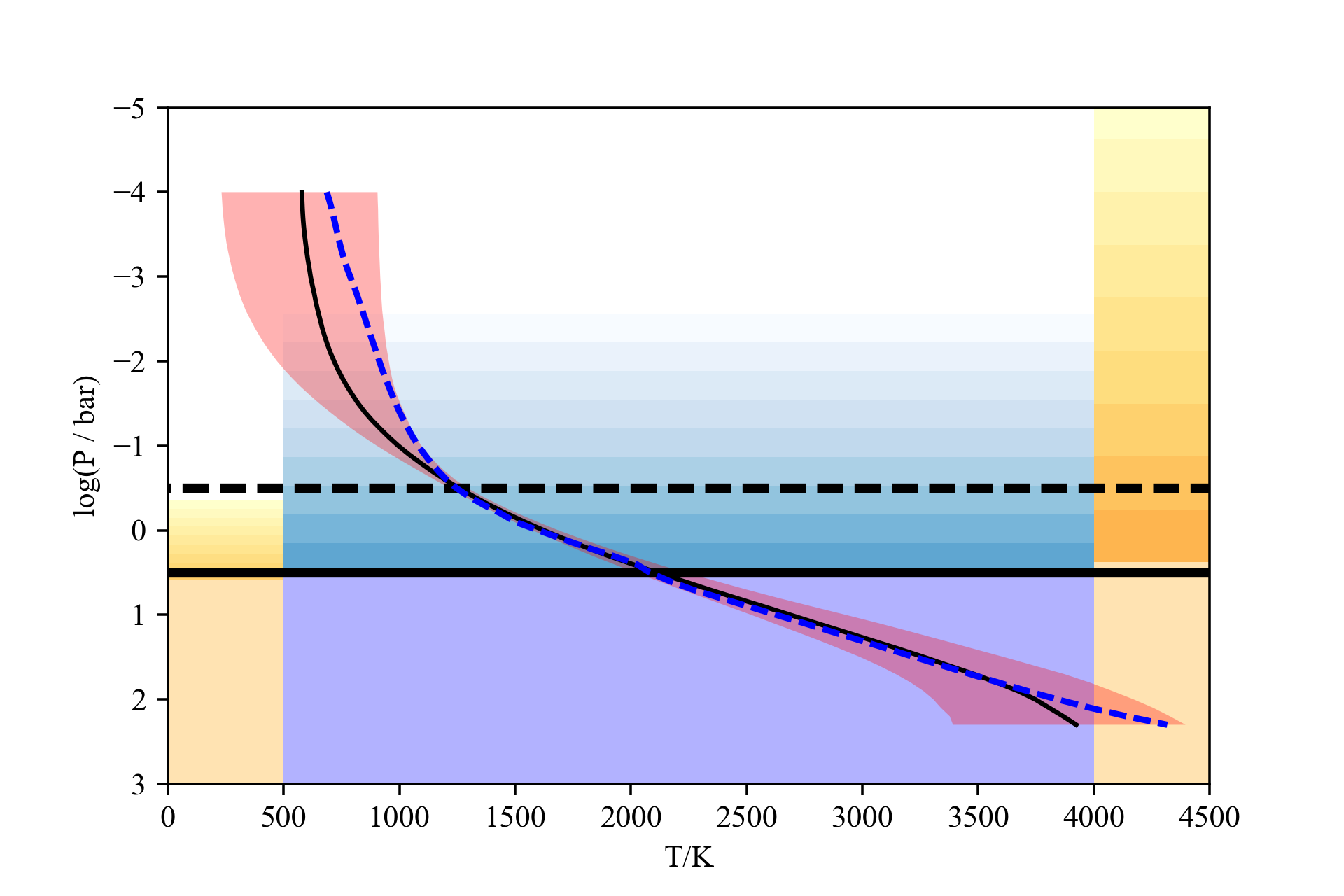}
\caption{Comparison of our retrieved thermal profile and cloud location with the input for our test case 2. Our median retrieved profile is indicated by a solid black line, and its 65\% confidence interval is denoted by red shading. The input profile is shown with a blue dashed line. The cloud deck pressure is indicated by a solid horizontal line, and its scale height is indicated with a horizontal dashed line. The median retrieved cloud is shaded blue at higher pressure than that of the cloud deck, and its scale height is indicated by graduated shading. The 68\% confidence range of the cloud is similarly indicated in yellow on either side of the plot.
\label{fig:test2_cld}}
\end{figure}

In Case 2 we adopt an optically thick grey cloud deck, whose optical depth passes unity at $\log P({\rm bars}) = 0.5$, and which has a scale height of 1~dex. The rest of the atmospheric parameters are as for Case 1.  Figure~\ref{fig:test2_cld} shows the retrieved profile and cloud location for the retrieval of this test case under the assumption of the optically thick cloud deck. As with the thin slab, the retrieved pressure at which the cloud optical depth reaches unity is located accurately. The vertical extent of the optically thin region of the cloud are not well constrained by the retrieval. The retrieved gas abundances in this case match the input values well, as does the retrieved profile. 

\begin{figure}
\hspace{-0.8cm}\includegraphics[width=290pt]{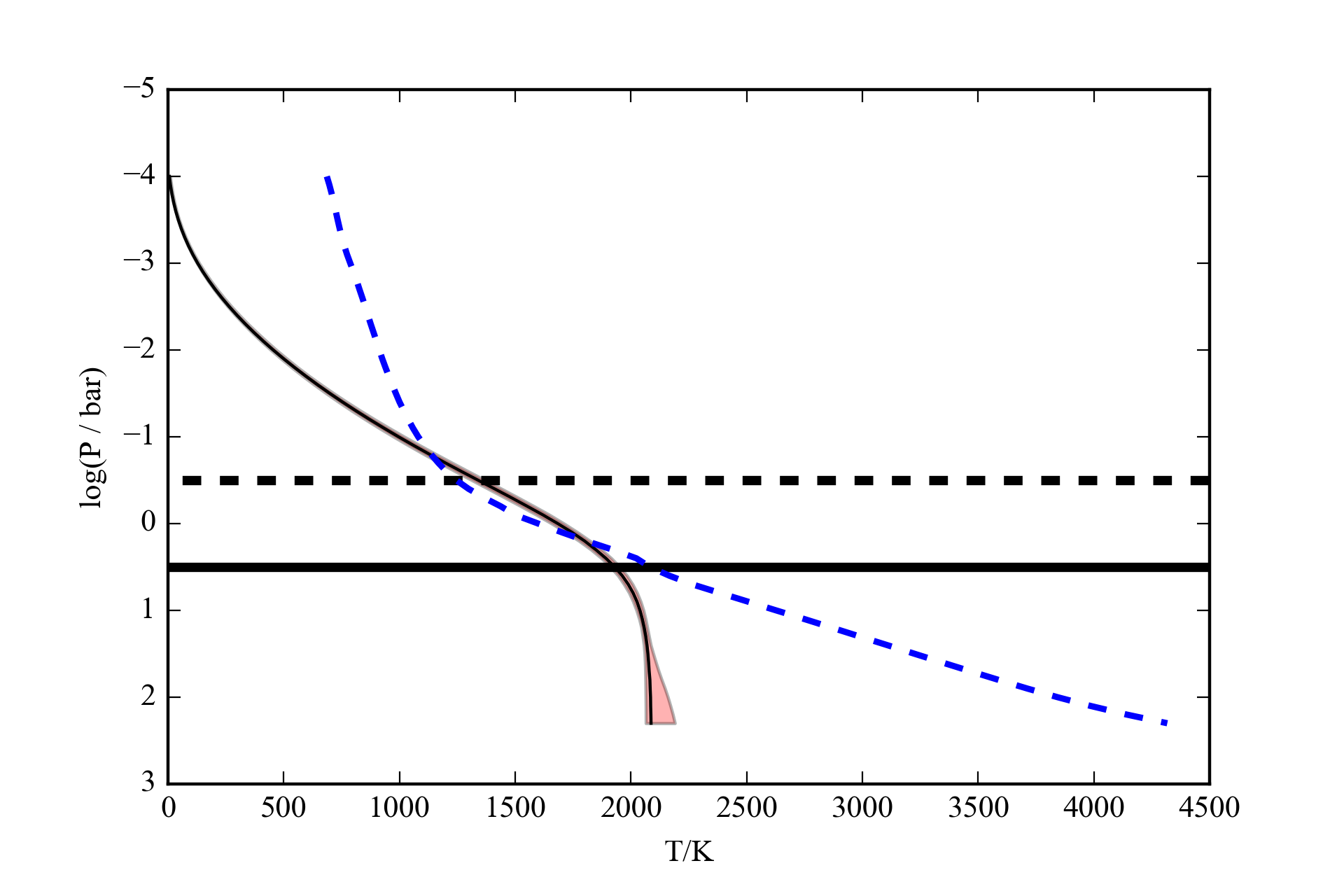}
\caption{Comparison of our retrieved thermal profile with the input for our test case 2, retrieved under a cloud-free assumption, following notation of Figure~\ref{fig:test2_cld}.  We have included the location of the input cloud deck as a horizontal solid black line, and the scale height over which its opacity drops off is indicated as a dashed black line.
\label{fig:test2_nc}}
\end{figure}

Under the cloud-free assumption, the vast majority of samples from the posterior distribution lie within a very narrow profile region (Figure~\ref{fig:test2_nc}), although inspection of a set of random draws from the posterior reveals several outlying clusters of profiles. In a more extreme version of the behaviour seen for the thin slab cloud in Case 1, the optically thick region of cloud is matched with a nearly isothermal profile, deviating strongly from the input profile. The profile deviates to near isothermal at the pressure that the cloud optical depth passes unity in the input spectrum model. 
This is driven by the need to truncate the flux in the bright opacity windows that are otherwise obscured by the cloud opacity. Like an opaque cloud, the isothermal profile similarly radiates as a single blackbody at the turnover temperature.
In this case the retrieved gas abundances are not as accurate as for the cloudy retrieval, and the retrieved gravity is well removed from the input value. The cloud-free case is unable to match the shape of the synthetic spectrum effectively, in particular it struggles with the bright $J$ band regions where it consistently overestimates the flux. Selecting between these cases using the BIC very strongly prefers the cloudy retrieval, with a $\Delta BIC = 62$.

\subsection{Summary of tests}

In both our test cases we were able to distinguish the cloudy and cloud-free scenarios using the BIC, although in Case 1 the power of this selection was considerably weaker than in Case 2, reflecting the lower impact of the $\tau = 1$ cloud compared to the extremely optically thick cloud deck used in Case 2. The tests reveal that we cannot expect to retrieve the vertical extent of clouds with any degree of precision. However, we are able to correctly retrieve the depth at which a cloud becomes optically thick. In the thin slab case, our ability to retrieve correct gas abundances is only weakly dependent on the cloud properties. In the cloud deck case the gas abundances differ strongly between the cloudy and cloud-free cases, and the cloudy retrieval successfully retrieved all parameters accurately. In both cases, the retrieved thermal profile depends strongly on the presence or otherwise of clouds in the retrieval model, with a shallow temperature gradient able to broadly mimic the influence of the cloud.  We also note that the parameterised profile is able to accurately fit and retrieve grid model thermal profiles in L~dwarf regime, and the profile retrieved by \citet{line2015} for G570D.

\section{First applications to field L dwarfs}
\label{sec:2m2224}
As our first application of this tool to the warm, cloudy, brown dwarf regime we attempt to retrieve the atmospheric properties of the L4 spectral template 2MASS~J05002100+0330501 \citep[2M0500+0330; ][]{reid2008} and the well-studied L4~dwarf 2MASSW~J2224438-015852 \citep[2MASS2224-0158; ][]{kirkpatrick2000}. We have selected these targets due to the availability excellent quality SNR $\sim$ 100,  0.7 -- 2.5 $\micron$ low-resolution Spex \citep{spex} spectroscopy, and their mid-L spectral types. The observed properties of our targets found in the literature are summarised in Table~\ref{tab:properties}.

\subsection{2M0500+0330}

2M0500+0330 is an L4 spectral template \citep{reid2008} at a distance of $13.54 \pm 0.35$~pc \citep{faherty2009}.  \citet{gagliuffi2014} found no evidence for unresolved binarity via spectral deconvolution, and \citet{gagne2015} find no evidence of youth or moving group membership.  \citet{filippazzo2015} estimated $T_{\rm eff} = 1793 \pm 72$ and $\log g = 5.2 \pm 0.19$  based on its bolometric luminosity and with reference to evolutionary models assuming an age range of 0.5 -- 10~Gyr.

\subsection{2M2224-0158}

This L4.5 dwarf lies a distance of $11.35 \pm 0.14$~pc \citep{faherty2012}. \citet{filippazzo2015} have estimated $T_{\rm eff} = 1646 \pm 0.71$K and $\log g = 5.18 \pm 0.22$ for this target using the same method and assumptions as for 2M0500+0330. It is somewhat redder than is typical for its spectral type, with $J-K = 2.1 \pm 0.03$. \citet{stephens2009} found that fitting grid models to the near-infrared spectrum of this object suggested a $T_{\rm eff}$ some 250~K cooler than implied by its luminosity.  \citet{cushing2006} presented a tentative detection of a silicate scattering feature in {\it Spitzer} IRS data for this target. This makes it an ideal test case to investigate the influence of silicate clouds on the emergent near-infrared spectrum. \citet{sorahana2014} used AKARI 3 -- 5$\mu$m spectroscopy to argue that chromospheric heating has resulted in an isothermal upper atmosphere ($\log P < -0.5$) with $T = 1445$K for 2M2224-0158. The notion of chromospheric heating is also consistent with the observation of H$\alpha$ emission in this target \citep{kirkpatrick2000}.

\begin{table*}
\begin{tabular}{c c c c }
\hline
Name & 2MASS0500+0330 & 2MASS2224-0158\\
$\alpha_{J2000}$ & 05:00:21.00 & 22:24:43.8$^1$ \\
$\delta_{J2000}$ & +03:30:50.18 & -01:58:52.14$^1$ \\
$J_{2MASS}$& $13.67 \pm 0.02^1$ &  $14.07 \pm 0.03^1$ \\
$H_{2MASS}$ & $12.68 \pm 0.02^1$ &  $12.82 \pm 0.03^1$ \\
$K_{2MASS}$ & $12.06 \pm 0.02^1$ & $12.02 \pm 0.02^1$\\
WISE-W1 & $11.51 \pm 0.02$  & $11.36 \pm 0.02$ \\
WISE-W2 & $11.25 \pm 0.02$ & $11.12 \pm 0.02$ \\
WISE-W3 & $10.96 \pm0.11$ & $10.65 \pm 0.09$ \\
Spectral type & L4 (opt)$^2$  / L4 (NIR)$^3$ & L3.5 (opt) / L4.5(NIR)$^4$ \\
Parallax & $73.85 \pm 1.98^5$ & $88.10 \pm 1.10^6$ \\
\hline
\end{tabular}
\caption{The observed properties of our retrieval targets. References: $^1$\citet{2mass}; $^2$ \citet{reid2008};$^3$ \citet{gagliuffi2014}; $^4$\citet{stephens2009}; $^5$\citet{dieterich2014}; $^6$ \citet{faherty2009}
\label{tab:properties}
}
\end{table*}

\subsection{Retrieval setup}
The set of gases retrieved were H$_{2}$O, CO, TiO, VO, CaH, CrH, FeH, Na and K. As in \citet{line2015}, the alkalis, were tied together as a single element in the state-vector assuming a Solar ratio taken from \citet{asplund2009}. We carried out retrievals under a variety of cloud assumptions that built up incrementally in complexity, and we assessed performed model selection based on the BIC.  Each retrieval was initialised with 16 walkers per parameter in a tight gaussian ball centred around the approximate Solar composition equilibrium chemistry values (see Section~\ref{subsec:comp}) for gas volume mixing ratios, wavelength shift between the data and the model,$\Delta \lambda = 0$, and for a scale factor consistent with $R \approx1.0 R_{jup}$,  but with flat distributions for gravity in the range $\log g = 4.5 - 5.5$, and covering the entire prior range for cloud properties and the tolerance parameter. 

We experimented with a variety of initialisations for the parameterised thermal profile. These were all based on fitted values of the $\alpha_{1}$, $\alpha_{2}$, $P_1$, $P_3$, $T_3$  parameters to grid model profiles from \citet{sm08}.  Our final initialisation was a gaussian distribution with standard deviations of $\pm 500$K about $T_3$ for the fit to a $T_{\rm eff} = 1700$K, $\log g = 5.0$ model, $\pm 0.2$dex distributions in $\log P$ for $P_1$ and $P_2$, and $\pm 0.1$ and $\pm 0.05$ in $\alpha_{1}$ and $\alpha_{2}$ about the same fitted values.  We also experimented with using grid model profiles in the range $T_{\rm eff} = 1500 - 1700$K, and with wider and narrower gaussian spreads, and with using flat distributions of values.  We found the choice amongst these had little impact on the final results. 

\subsection{Model selection}

The BICs for our retrievals using different cloud models are shown in Table~\ref{tab:bic}.  The hierarchy of preferred models is different for our two targets, but in both cases the optically thick, power-law cloud deck is strongly preferred over all other models.

\begin{table}
\begin{tabular}{c c c c}
\hline
Cloud case & $N_{param}$ & \multicolumn{2}{ c }{$\Delta BIC$} \\
& & 2M0500 & 2M2224 \\
\hline
Cloud free & 17 &  +11 & +50 \\
Grey cloud deck & 20 & +11 & +40 \\
Power-law cloud deck & 21 & 0 & 0 \\
Slab grey cloud & 21 & +20  & +48  \\
Slab power-law cloud & 22 & +13 & +36 \\
Two clouds: slab + deeper deck & 26 & +22 & +26  \\
\hline
\end{tabular}
\caption{$\Delta$BIC values for our field dwarf retrievals. Our preferred power-law cloud deck model is taken as the reference point. 
\label{tab:bic}}
\end{table}

\subsection{The retrieved spectra}

\begin{figure}
\hspace{-0.8cm}\includegraphics[width=285pt]{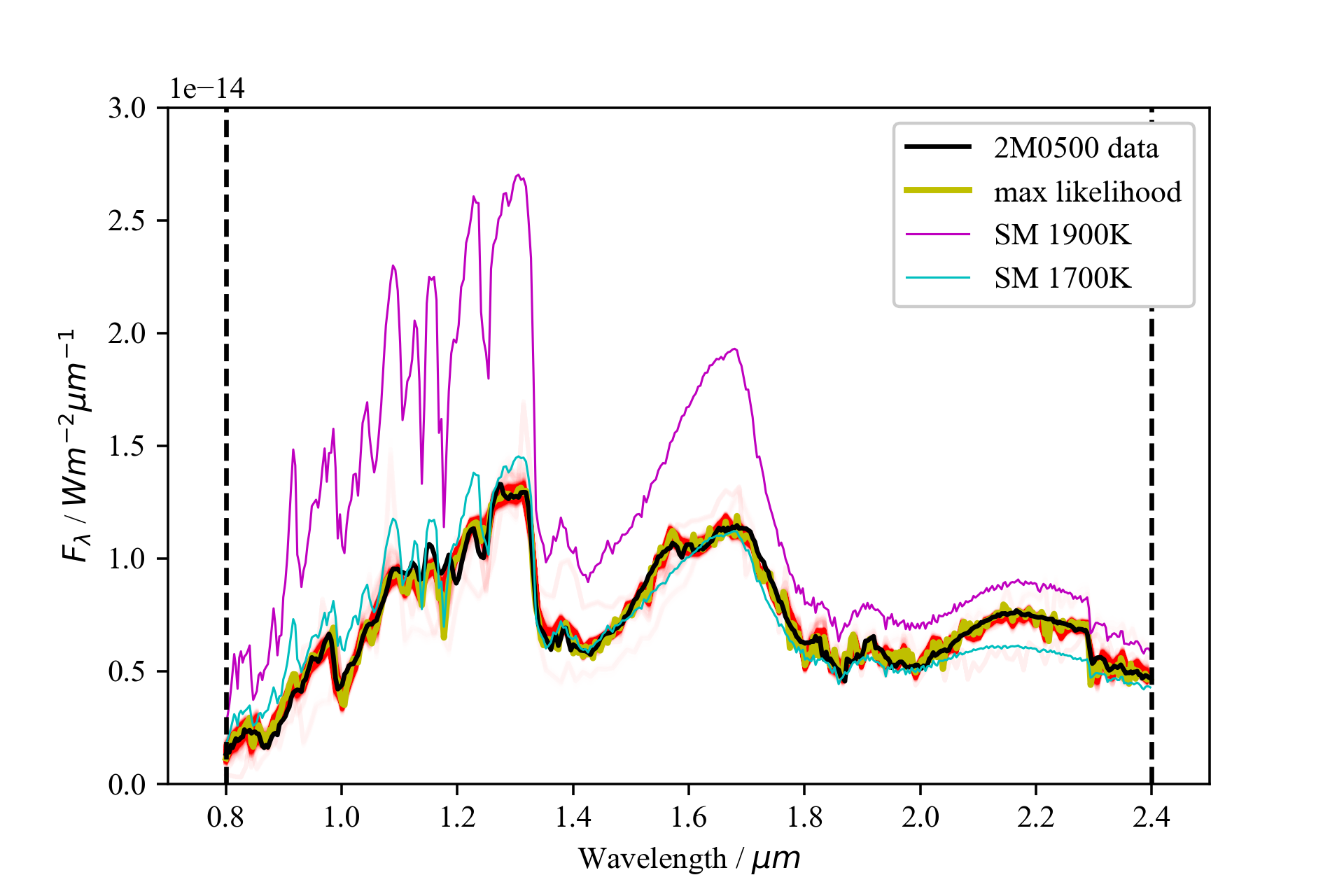}
\caption{ The 2M0500+0330 forward model output spectra for 5000 random draws from our final 2000 iterations of the {\sc emcee} sampler, comprising 32000 independent samples, shown in red. The maximum likelihood case is plotted in yellow, and the Spex spectrum for our target is plotted in black \citep{gagliuffi2014}. Also plotted are the Saumon \& Marley grid model $f_{sed} = 2$ predictions for spectra for $\log g = 5.0$ and $T_{\rm eff} = 1700$K and 1900~K, which bracket the $T_{\rm eff}$ estimated by \citet{filippazzo2015}.
\label{fig:2m0500thickspec}}
\end{figure}

\begin{figure}
\hspace{-0.8cm}\includegraphics[width=285pt]{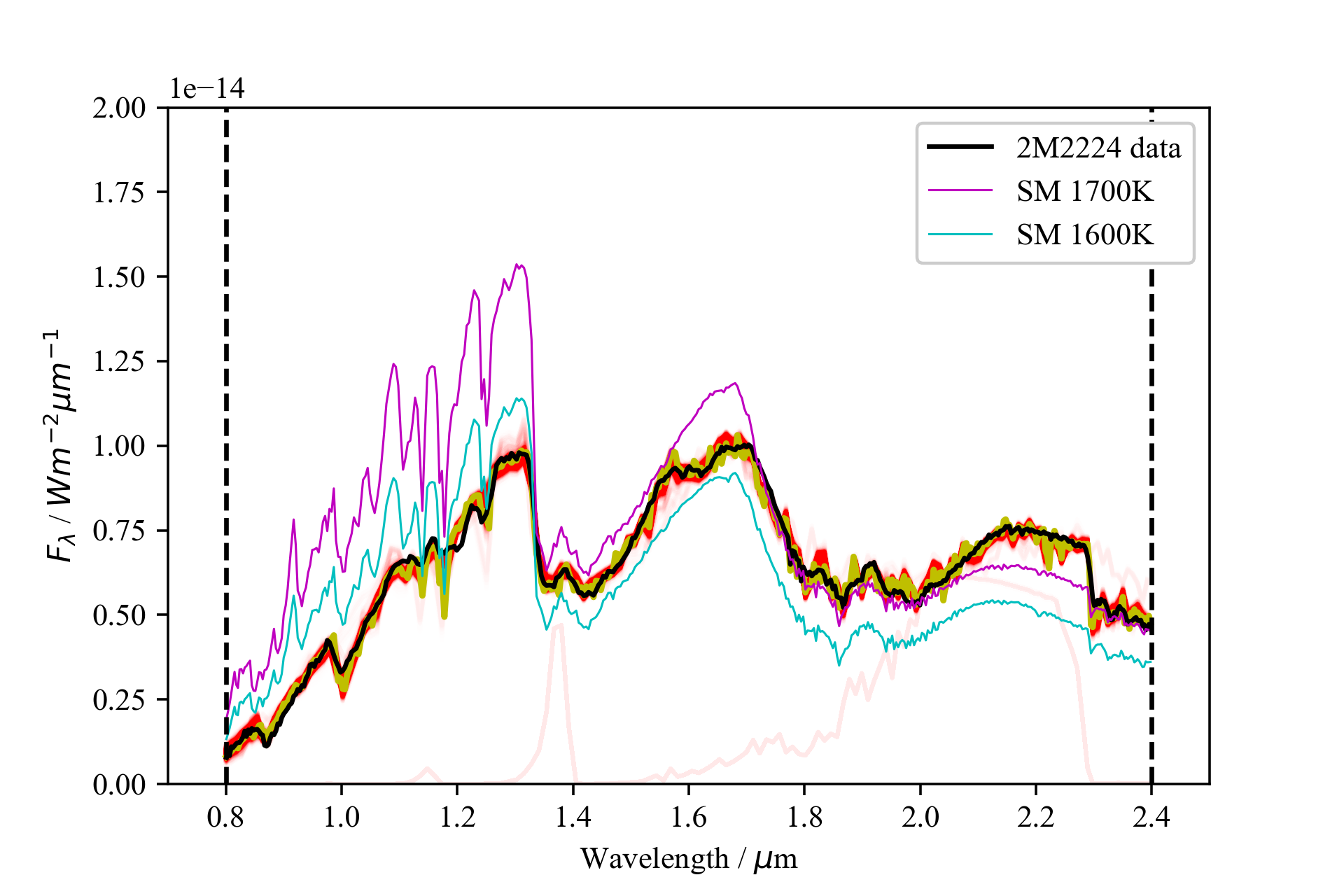}
\caption{ The 2M2224+0158 forward model output spectra for 5000 random draws from our final 2000 iterations of the {\sc emcee} sampler, comprising 32000 independent samples, shown in red. The maximum likelihood case is plotted in yellow, and the Spex spectrum for our target is plotted in black \citep{burgasser2010a}. Also plotted are the Saumon \& Marley grid model $f_{sed} = 2$ predictions for spectra for $\log g = 5.0$ and $T_{\rm eff} = 1600$K and 1700~K, which bracket the $T_{\rm eff}$ estimated by \citet{filippazzo2015}.
\label{fig:2m2224thickspec}}
\end{figure}

\begin{figure}
\hspace{-0.8cm}\includegraphics[width=285pt]{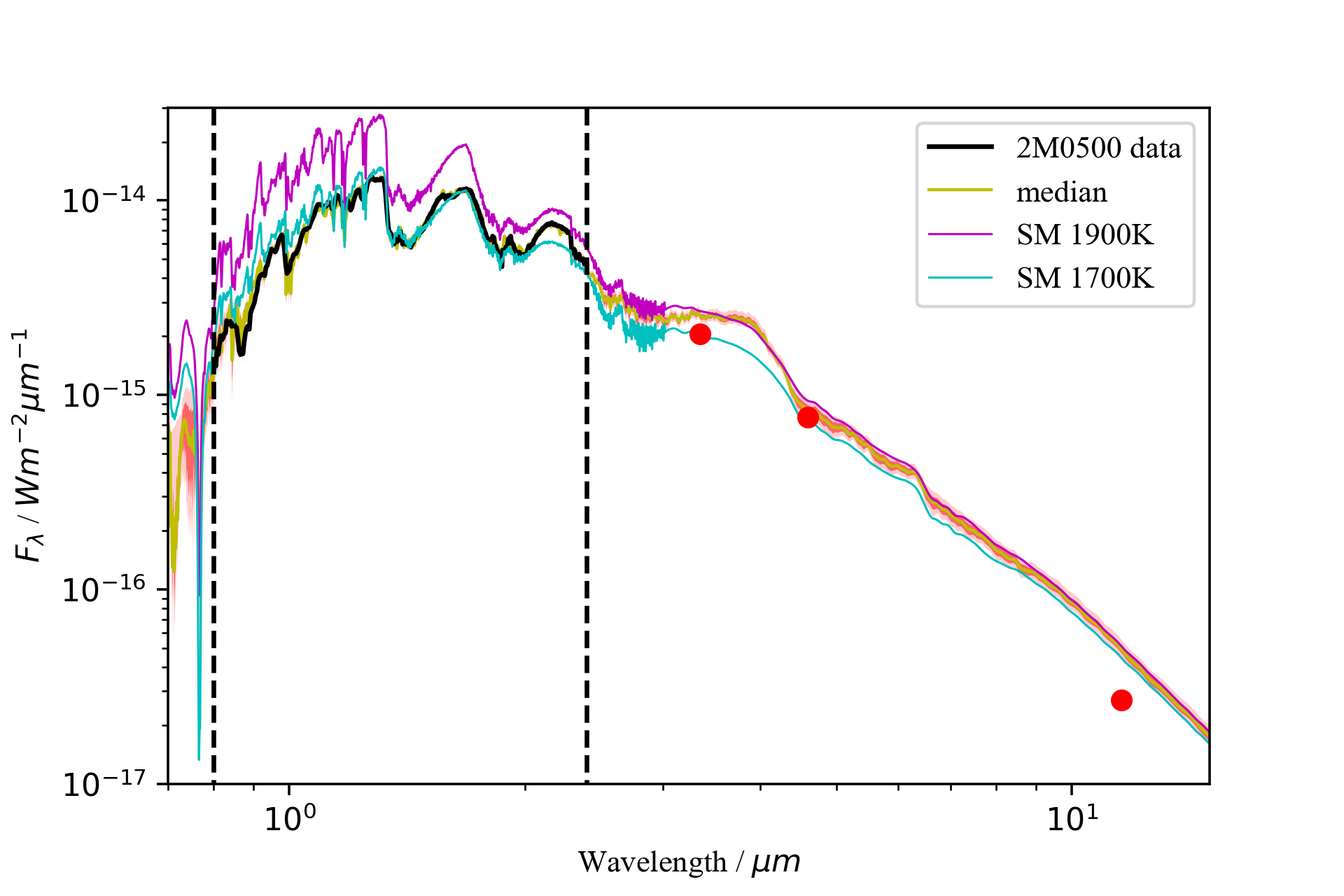}
\caption{The extension of our median spectrum for 2M0500+0330 to longer wavelengths compared to WISE photometric fluxes (red filled circles) and the grid models shown in Figure~\ref{fig:2m0500thickspec}. The 1$\sigma$ and 2$\sigma$ interval for the spectrum is indicated by red and pink shading. The uncertainties on the WISE points are comparable to the symbol size.  
\label{fig:longspec1}}
\end{figure}

\begin{figure}
\hspace{-0.8cm}\includegraphics[width=285pt]{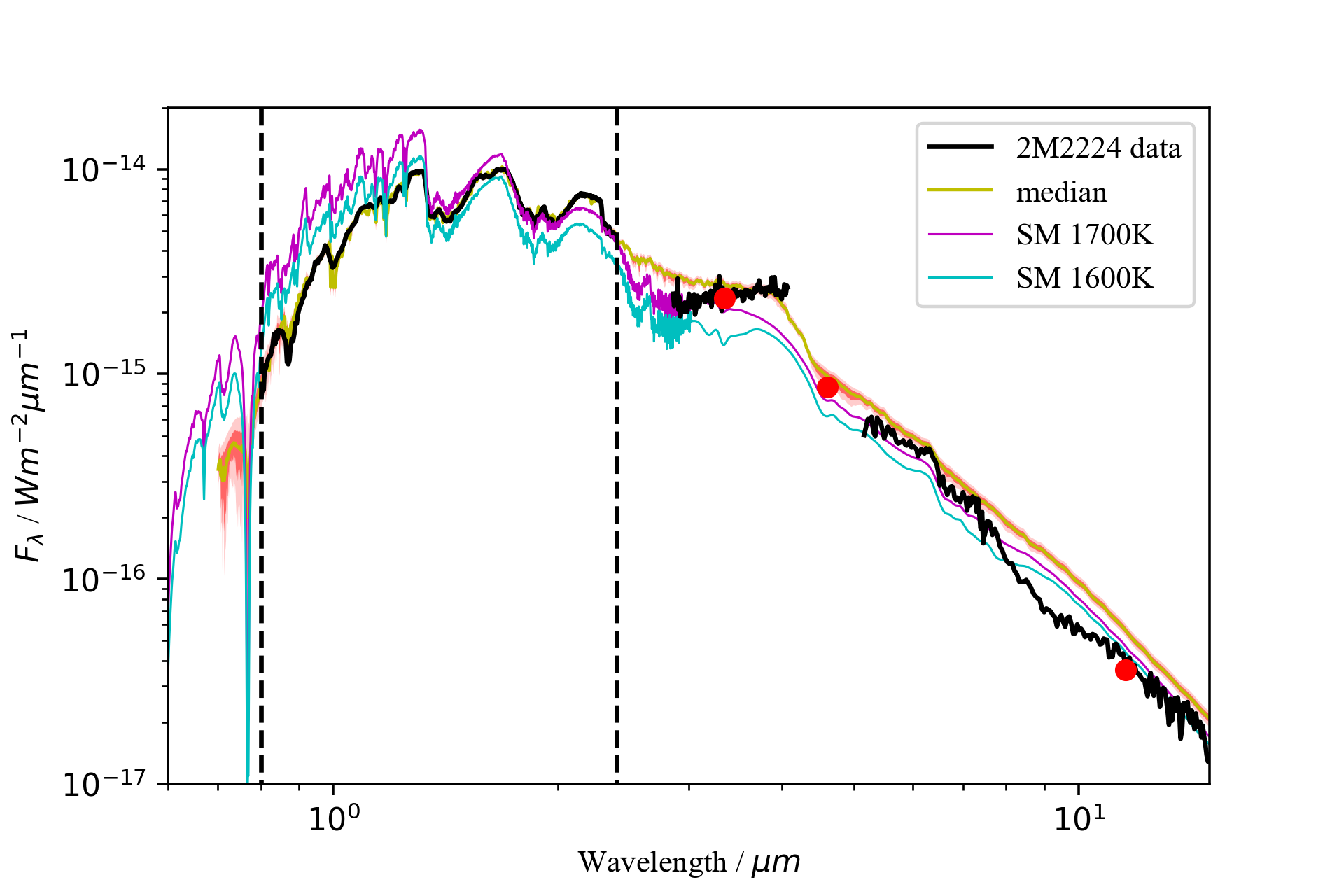}
\caption{The extension of our median spectrum for 2M2224-0158 to longer wavelengths compared to WISE photometric fluxes (red filled circles) and the grid models shown in Figure~\ref{fig:2m2224thickspec}. The 1$\sigma$ and 2$\sigma$ interval for the spectrum is indicated by red and pink shading. The uncertainties on the WISE points are comparable to the symbol size.  Also plotted are the longer wavelength spectra for this target from \citet{cushing2006}.
\label{fig:longspec2}}
\end{figure}

In Figures~\ref{fig:2m0500thickspec} and~\ref{fig:2m2224thickspec} we compare the forward model spectra for our retrieved parameters to the observed spectra for our targets. The match is extremely good, particularly compared to the cloudy ($f_{sed} = 2$) Marley \& Saumon grid models for the gravity and $T_{\rm eff}$ estimated by \citet{filippazzo2015}. It is worth noting that while the two grid model spectra effectively bracket our retrieval spectrum for 2M2224, they both have overall slopes that are too blue between the $J$ and $K$ band regions (1.2 -- 2.5 $\mu$m) compared to the observed spectrum. In Figures~\ref{fig:longspec1} and~\ref{fig:longspec2},  we extend our maximum likelihood spectrum to longer wavelengths to allow comparison with our targets' WISE photometry \citet{wise}. Our spectra match the WISE-W2  channel flux near 4.5$\mu$m well, but in both cases our flux is somewhat overestimated at the shorter wavelength WISE-W1 point. This is not necessarily surprising since the WISE-W1 band was designed to be sensitive to the depth of the \meth~$\nu_{3}$ fundamental band at 3.3$\mu$m, and we have not included~\meth in our forward model for this case due to its weak impact in the near-infrared at these temperatures. Although methane is typically associated with objects in the T~dwarf regime, it has been detected L~dwarfs as early in type as L5 due to its much stronger feature at 3.3$\mu$m \citep[e.g. ][]{noll2000,stephens2009}.

\subsection{The retrieved profiles}
\label{sec:prof}

\begin{figure}
\hspace{-0.8cm}\includegraphics[width=290pt]{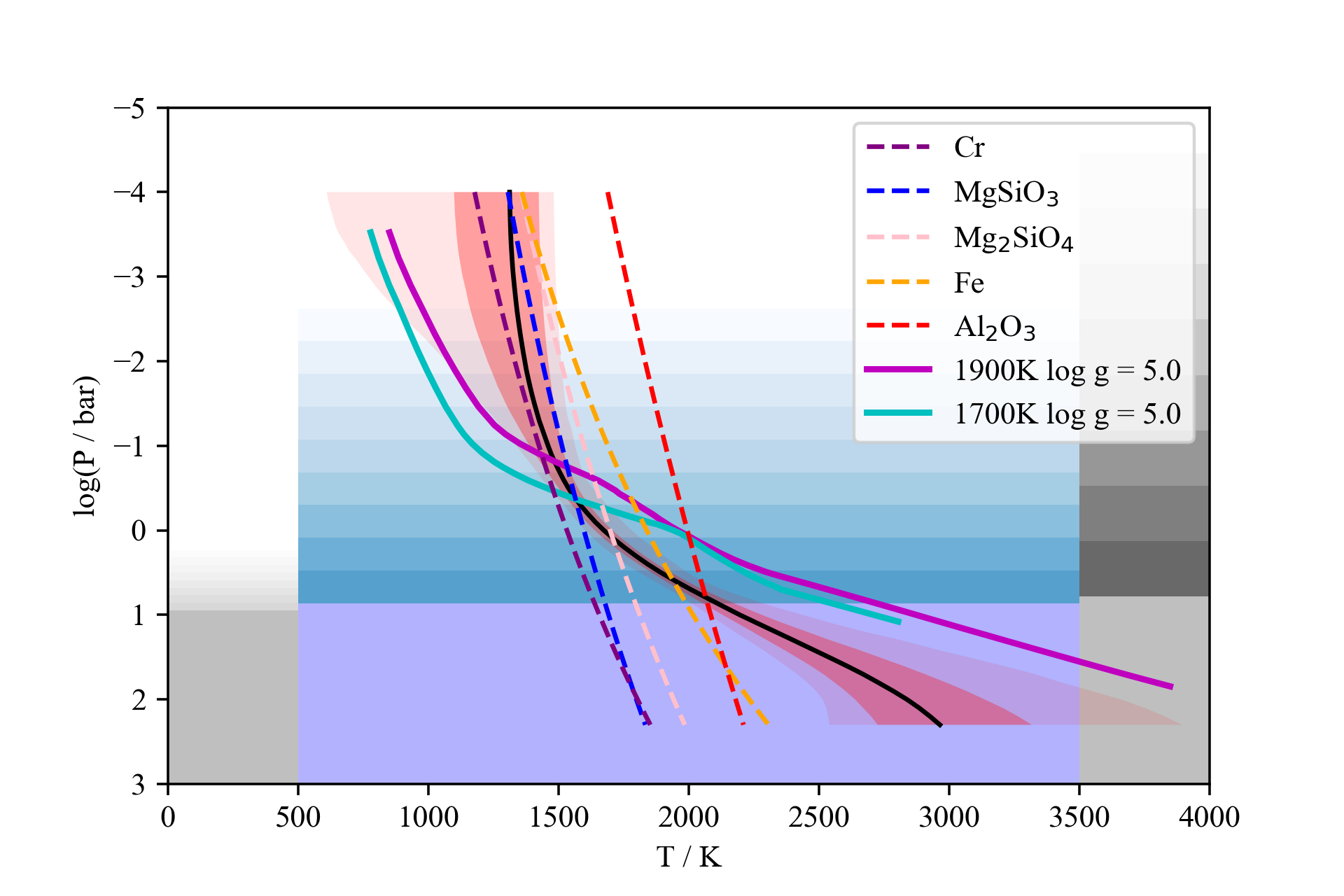}
\caption{Our retrieved thermal profile and cloud location for our preferred optically thick cloud deck model for 2M0500-0330. The median case thermal profile is shown as a solid black line, with the 1$\sigma$  and 2$\sigma$ intervals are indicated with red and pink shading. The central shaded panel indicates the median cloud location, whilst the edge panels show the $\pm 1\sigma$ cloud profiles from the retrieval. The cloud top, where the cloud optical depth passes unity, is indicated by extent of the solid shading, whilst the scale height of the cloud above the top is shown with graded shading. Also plotted are condensation curves for likely cloud species, and the forward model grid (T,P) profiles for $f_{sed} = 2$, $\log g = 5.0$,  for $T{\rm eff} =  1700$K and 1900~K, which bracket the $T_{\rm eff}$ estimated by \citet{filippazzo2015}, and that inferred from this retrieval model.
\label{fig:2m0500thickprof}}
\end{figure}

\begin{figure}
\hspace{-0.8cm}\includegraphics[width=290pt]{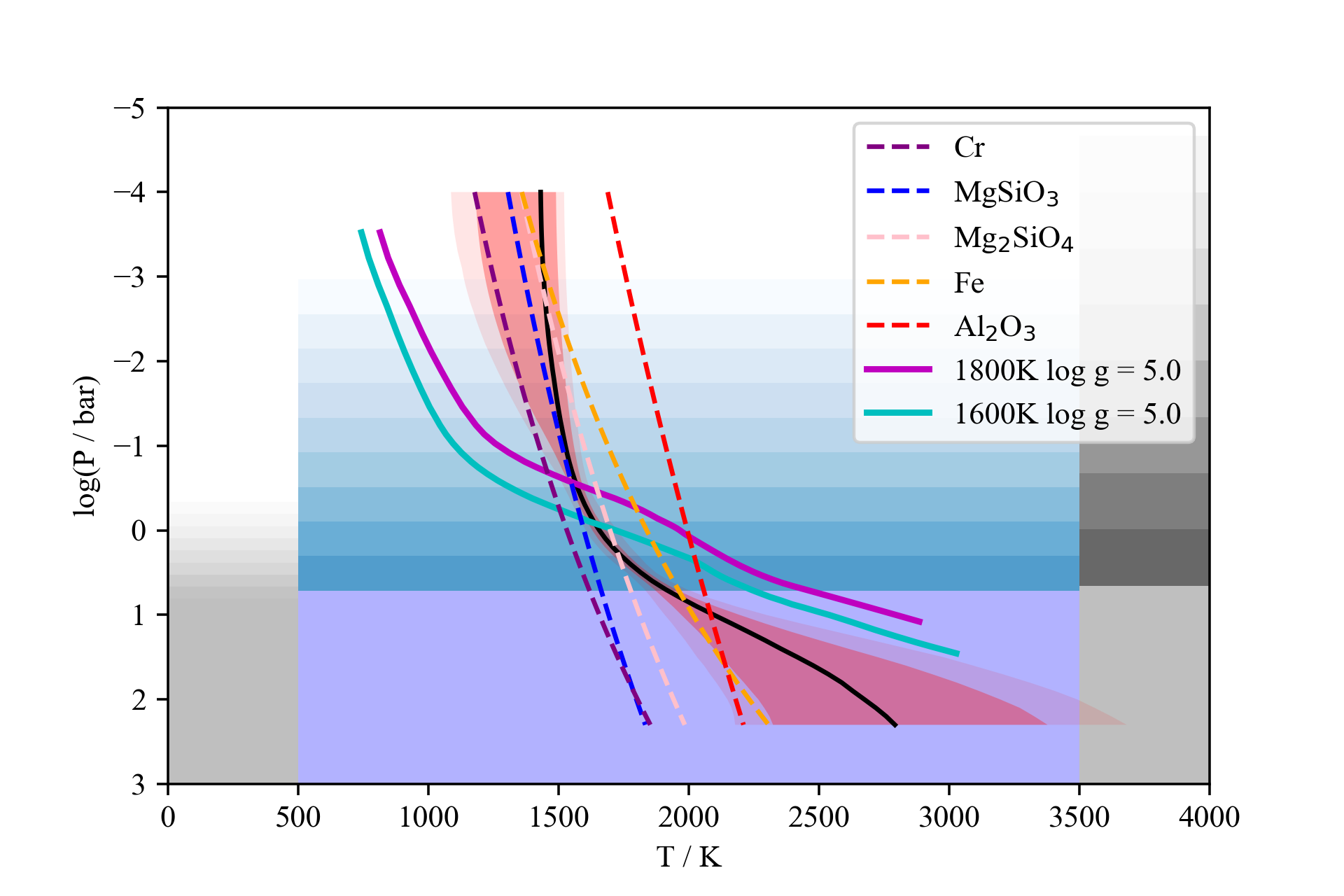}
\caption{Our retrieved thermal profile and cloud location for our preferred optically thick cloud deck model for 2M2224-0158. The median case thermal profile is shown as a solid black line, with the 1$\sigma$  and 2$\sigma$ intervals are indicated with red and pink shading. The central shaded panel indicates the median cloud location, whilst the edge panels show the 1$\sigma$ boundary cases. The cloud top, where the cloud optical depth passes unity, is indicated by extent of the solid shading, whilst the scale height of the cloud above the top is indicated with graded shading. Also plotted are condensation curves for likely cloud species, and the forward model grid (T,P) profiles for $f_{sed} = 2$, $\log g = 5.0$,  for $T{\rm eff} =  1600$K and 1800~K, which which bracket the $T_{\rm eff}$ estimated by \citet{filippazzo2015}, and that inferred from this retrieval model.
\label{fig:2m2224thickprof}}
\end{figure}

Figure~\ref{fig:2m0500thickprof} and~\ref{fig:2m2224thickprof} shows our retrieved thermal profiles and the location of the cloud deck in our preferred retrievals. Our profiles differs significantly from the corresponding forward grid model, being significantly cooler in the region between 1~and~10~bar and by being nearly isothermal at pressures lower than 0.1 bar. The first of these likely goes some way to explaining the redder slopes in our retrieval spectra compared to the grid models, and the difficulty that the grid models have in correctly predicting the colours of L~dwarfs as a function of spectral type. Although our profile shape may be somewhat surprising in the context of model grid predictions, we note that its clearly driven by the data, and is well removed from the initialisation values that were based on these same grid models. The deep profiles are also close to parallel with the grid model predictions, and so do not appear to be unphysical. The upper atmosphere temperature is also consistent with the results of \citet{sorahana2014} for 2M2224-0158, and suggests that a heat source for the radiative equilibrium region of the atmosphere may be important in one or both of these targets.

It is important to note that the deep thermal profiles (at pressures higher than 10 bar) are really the extrapolation of the profile gradient at lower pressures, since there is little contribution to the observed flux from beneath the cloud deck. This is reflected in the widening confidence interval with increasing pressure.

At first glance, there is some similarity to the cloud-free retrieval of Test Case~1 that might lead one to attribute our profile differences to an omitted cloud: at lower pressure than the omitted cloud in Test Case 1, the retrieved profile was warmer than the input profile, while beneath the omitted cloud it was cooler.  However, there are two points that argue against this explanation for the disagreement between our retrieved profiles with the grid models'. Firstly, we have performed a retrieval on these targets using a two cloud model that allows for a thin slab cloud lying above the main cloud deck (which was rejected during model selection). This scenario did not retrieve a significantly different profile, and the additional cloud was placed within the deep cloud deck. Secondly, the deep cloud deck in the preferred model also incorporates an optically thin extension to lower pressures, the extent of which is not strongly constrained. We found in the test cases that the vertical extent of the cloud has little effect on the profile or other retrieved properties, although the $\tau_{cl} \geq 1$ level is important. So, whilst it remains a possibility that missing low-pressure cloud opacity may impact our retrieved profile, it would require sub-structure in the clouds' vertical distribution to be important in a way that so far appears unlikely.

\subsection{Composition, $T_{\rm eff}$ and gravity}
\label{subsec:comp}

\begin{figure*}
\hspace{-1.0cm}
\includegraphics[width=550pt]{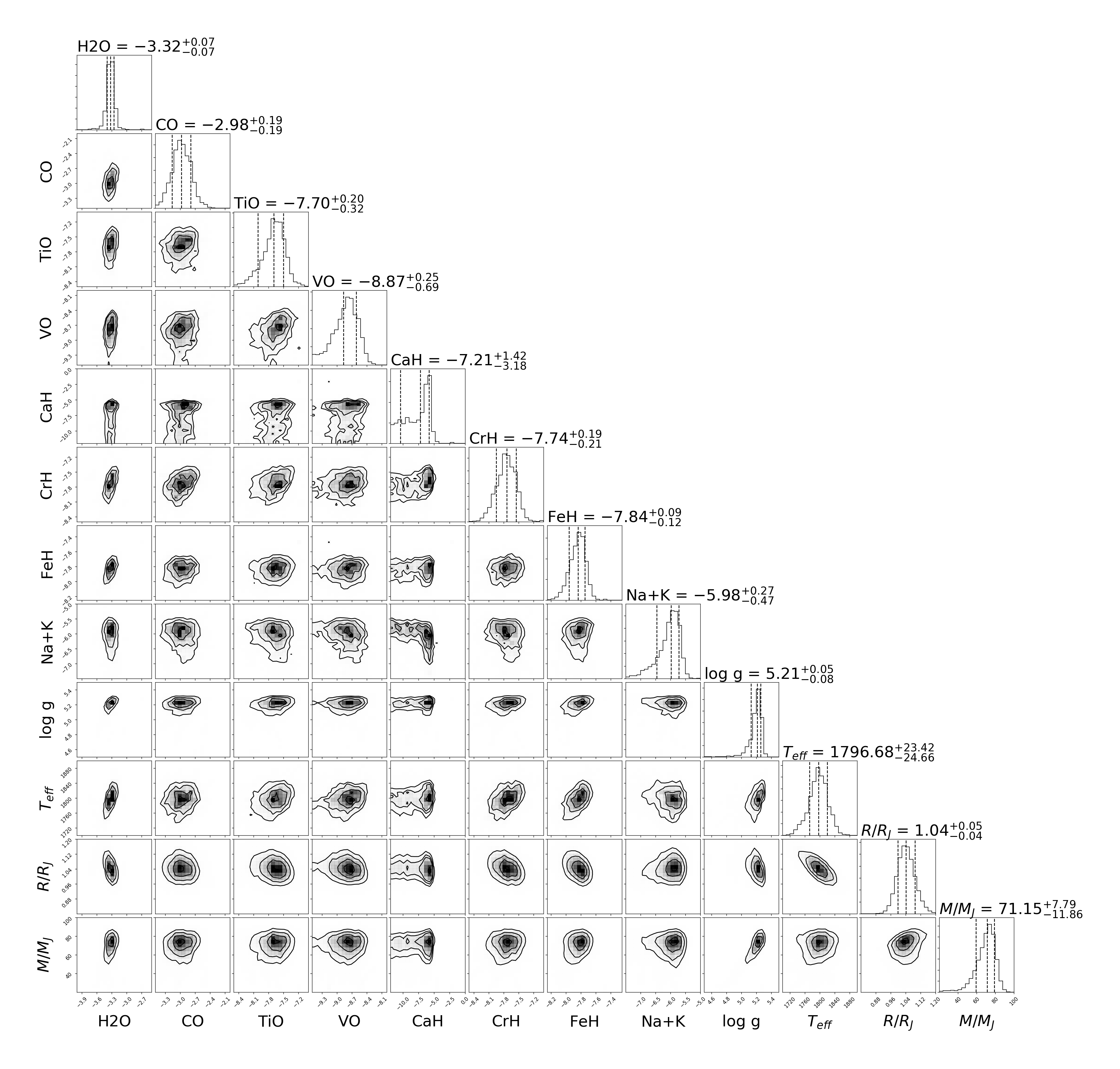}
\caption{Posterior probability distributions for some our retrieved parameters in the selected cloud deck case for 2M0500+0330. Also shown are $M$, $R$ and $T_{\rm eff}$ which are not directly retrieved, but are inferred from our retrieved parameters $R^2/D^2$, $\log g$ and predicted spectra. 
\label{fig:2m0500post}}
\end{figure*}

\begin{figure*}
\hspace{-1.0cm}
\includegraphics[width=550pt]{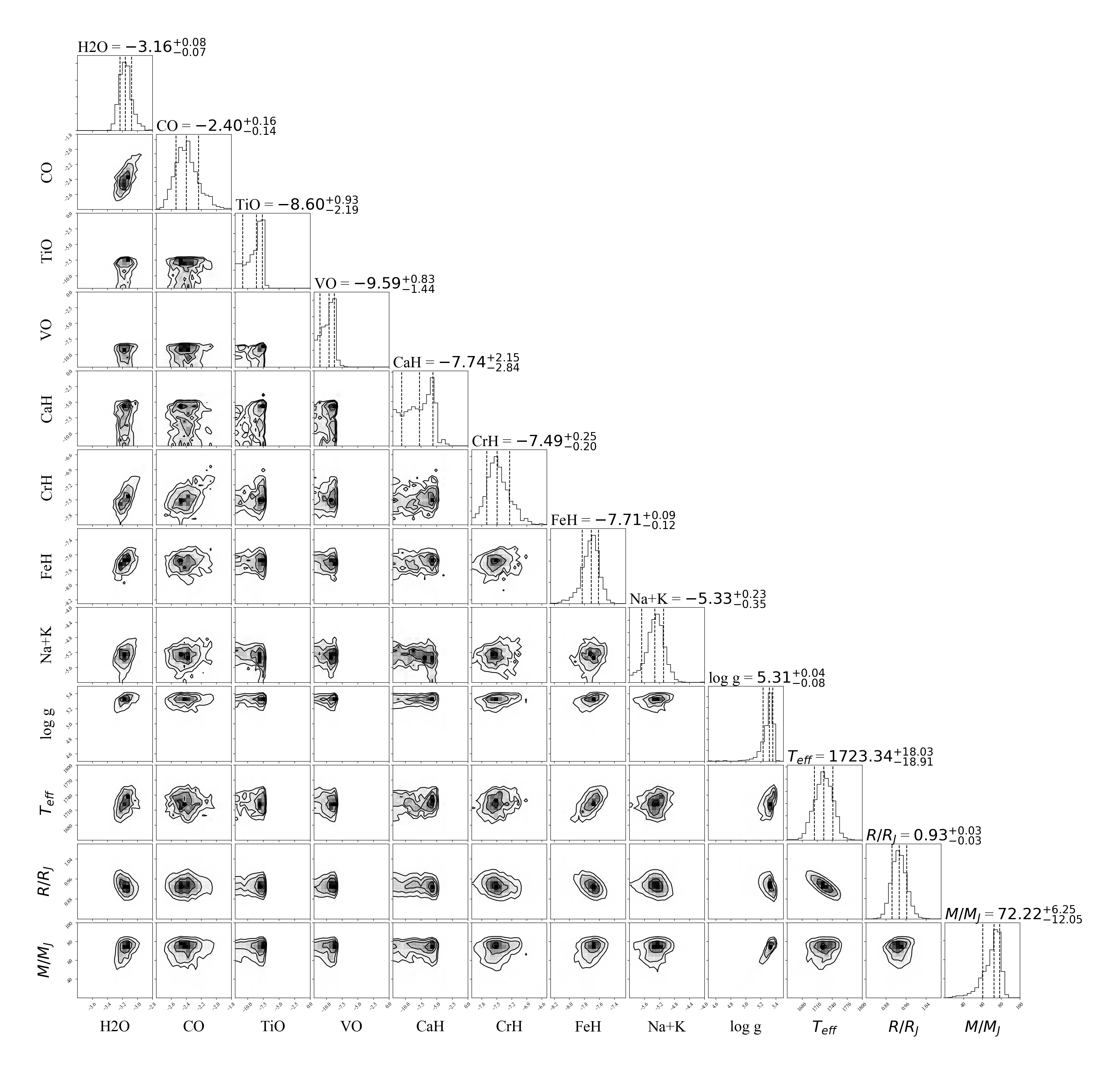}
\caption{Posterior probability distributions for some of our retrieved parameters in the selected cloud deck case for 2M2224-0158. Also shown are $M$, $R$ and $T_{\rm eff}$ which are not directly retrieved, but are inferred from our retrieved parameters $R^2/D^2$, $\log g$ and predicted spectra. 
\label{fig:2m2224thickPOST}}
\end{figure*}

Figures~\ref{fig:2m0500post} and~\ref{fig:2m2224thickPOST} shows the posterior probability distributions for our retrieved gas abundances, along with gravity and the derived quantities $M$, $R$ and $T_{\rm eff}$. The radius and mass are calculated from the retrieved $R^2/D^2$ scaling factor, trigonometric parallax and gravity. We note that despite our prior requirement that $M < 80 M_{Jup}$, our distributions in Figures~\ref{fig:2m0500post} and~\ref{fig:2m2224thickPOST} spread beyond this value. This is because our prior is applied during the retrieval assuming zero error on the distance to the target, whereas the subsequent evaluation of mass and radius posteriors incorporates the uncertainty in the trigonometric parallaxes. 
The $T_{\rm eff}$ is calculated using the derived radius and integrating the flux in our forward model output spectra between 0.6 and 20$\micron$. Our derived $T_{\rm eff}$ and gravities are in good agreement with those found by \citet{filippazzo2015}. For 2M0500+0330 we find $T_{\rm eff} = 1796^{+23}_{-25}$~K and $\log g  = 5.21^{+0.05}_{-0.08}$  (c.f. $T_{\rm eff} = 1793 \pm 72$~K and $\log g = 5.19 \pm 0.19$ ). For 2M2224-0158 our retrieved $T_{\rm eff} = 1723^{+18}_{-19}$~K and $\log g = 5.31^{+0.04}_{-0.08}$ are marginally higher, but consistent with the  Filipazzo values ( $T_{\rm eff} = 1646 \pm 71$~K; $\log g = 5.18 \pm 0.22$), which likely reflects the overestimated flux our model displays in the WISE-W1 region.  Our retrieved gravities, masses, radii and $T_{\rm eff}$ are also also all consistent with evolutionary model predictions for a Solar metallicity brown dwarf with an age in the 2 -- 10~Gyr range \citep{sm08,baraffe03,chabrier2000,baraffe2002}.

\begin{figure*}
\hspace{-0.8cm}
\includegraphics[width=350pt]{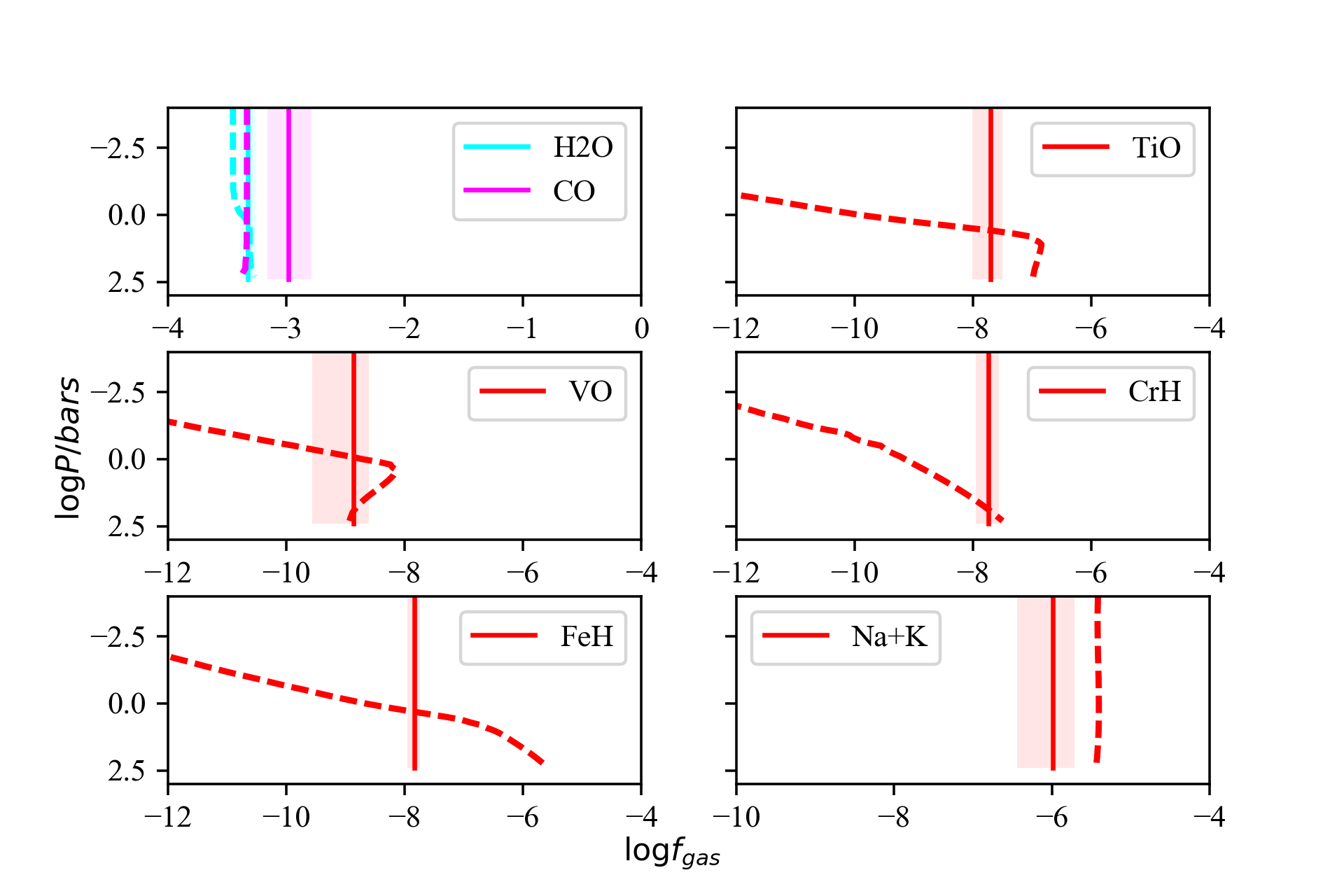}
\caption{Retrieved abundances for 2M0500+0330 and predicted thermochemical equilibrium abundances for our retrieved profile. 
\label{fig:2m0500abund}}
\end{figure*}

\begin{figure*}
\hspace{-0.8cm}
\includegraphics[width=350pt]{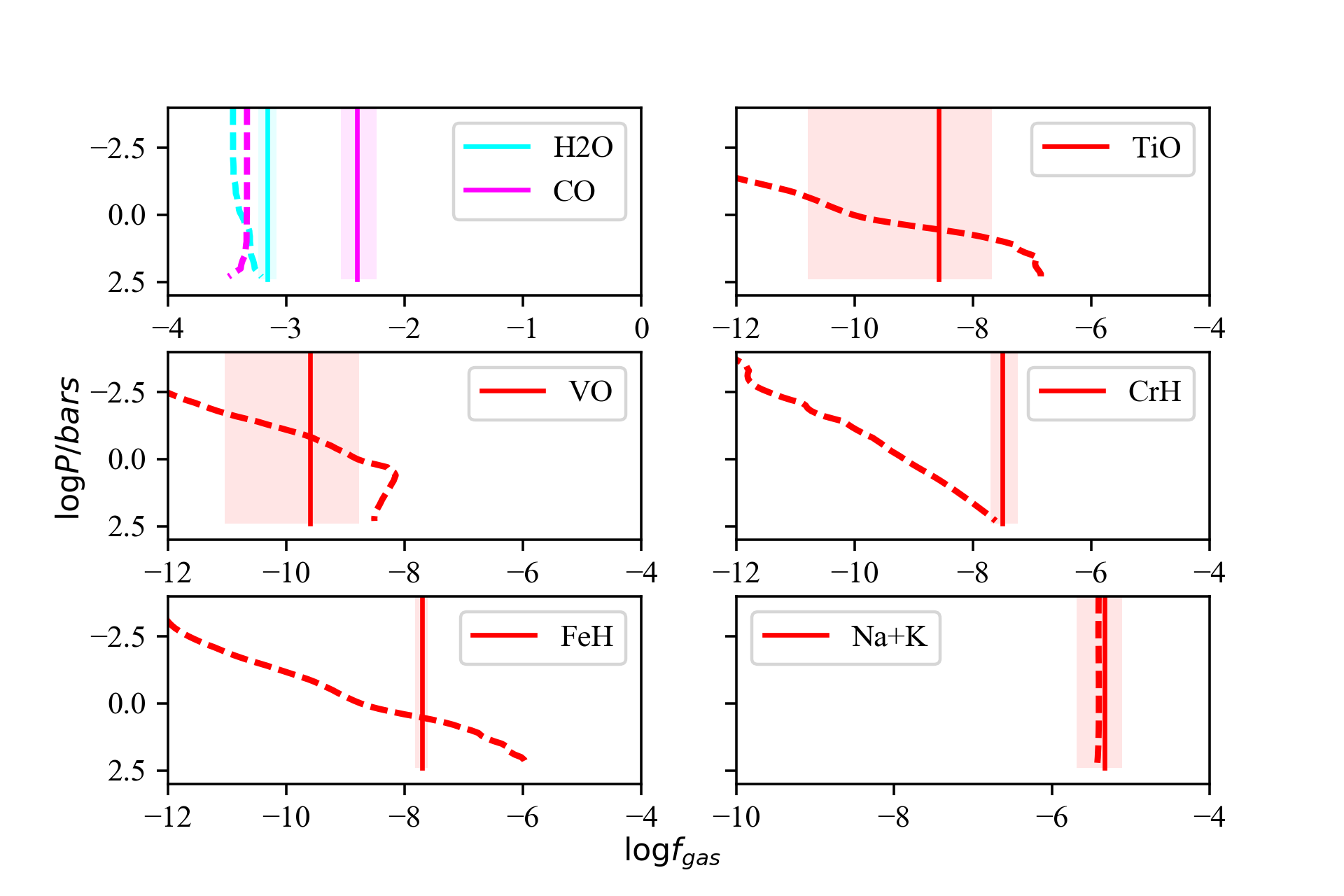}
\caption{Retrieved abundances for 2M2224-0158 and predicted thermochemical equilibrium abundances for our retrieved profile. 
\label{fig:2m2224abund}}
\end{figure*}

To place our retrieved gas abundances in context we compare them to modelled chemical equilibrium abundances in Figures~\ref{fig:2m0500abund} and~\ref{fig:2m2224abund}. 
The chemical equilibrium grids were calculated using the NASA Gibbs minimization CEA code \citep[see  ][]{mcbride1994}, based upon prior thermochemical models \citep{fegley1994,fegley1996,lodders1999,lodders2002,lodders2010,lodders2002b,lodders2006,visscher2006,visscher2010a,visscher2012,moses2012,moses2013} and recently utilized to explore gas and condensate chemistry over a range of substellar atmospheric conditions \citep{morley2012,morley2013,skemer2016,kataria2016,wakeford2016}.
The chemical grids are used to determine the equilibrium abundances of atmospheric species over a wide range of atmospheric pressures (from 1 microbar to 300 bar) and temperatures (300 to 4000 K), and in this case we assume Solar metallicity and C/O ratio.


The predicted abundance of several species varies by up to four order of magnitude across our atmospheric profile. This is in sharp contrast to the simple gas profiles seen in the T dwarf regime \citep[e.g.][]{line2015}.  This represents a significant challenge for our retrieval, which for expediency assumes a uniform-with-pressure volume mixing ratio for all gases. Investigating alternative methods for modelling the variation of abundance with altitude for gases in the L~dwarf regime will be investigated in a future work.  

Although some of our retrieved abundances are comparable to the predicted abundance in the region of the atmosphere where most flux originates (1 -- 10 bar), there is a striking disagreement between our retrieved CO abundance and the predicted value in both cases, with our retrieval finding a factor $\sim10$ higher CO abundance than predicted in the case of 2M2224-0158. In addition, for 2M0500+0330, our retrieved alkali abundance is 50\% smaller than predicted, and our CrH abundances in both cases appear overestimated. Without external constraints on the ``true" composition of our targets, it is difficult to interpret our retrieved compositions in any robust way.  

It is important to note that neither the BT Settl models \citep{btsettlCS16} nor the \citet{sm08} models are able to accurately reproduce both the near-infrared spectral shape and the depth of the CO band head. Indeed, both sets of models significantly under-predict the depth of this feature at Solar metallicity.  
However, given our retrieved profile's deviation from the grid model predictions, we first investigate if a steeper thermal profile in the upper atmosphere could improve the fit to the CO band with a lower abundance.   

We have performed retrievals for 2M2224-0158 in which we fix the profile as one of the grid profiles for the range $T_{\rm eff} = 1500 - 1700$K.  All but the $T_{\rm eff} = 1500$K case resulted in similarly high CO abundances, but with poor fits to the data. The $T_{\rm eff} = 1500$K case returned a CO abundance close to the predicted value and a with a good fit to the depth of the CO band. However, elsewhere it provided a significantly poorer fit to the data than our best retrieval, with substantial deviations in the $J$ band peak, the blue shoulder of the $H$ band, and in the shape of the $K$ band bump.  

We have also tested for a correlation between the slope of the thermal profile at pressures smaller than 0.1 bar and the CO mixing ratio in the final 10000 iterations of our {\sc emcee} chain (total 3,360,000 samples). We find that although there is a set of cases with thermal profiles with a steeper low-pressure gradient and a CO abundance near the grid-predicted value, these cases have much lower likelihood, apparently driven by their poor fit to the CO bandhead. We thus conclude that the thermal profile (and specifically its shallow gradient at pressures smaller than 0.1 bar) is unlikely to be the cause of our apparently anomalous CO abundance. We also note that our choice of cloud model does not strongly influence our retrieved CO fraction, although the cloud-free model requires moderately higher CO abundance. The retrieved CO fraction is also not correlated with cloud thickness or cloud scale height.

 The remaining relatively-simple origin of both our anomalous CO abundance and the poor performance of the grid models in this area is an issue with the CO cross-sections used by both the grid models and our retrieval model.

\subsection{Cloud properties}
\label{subsec:cloudret}

Our retrieval result allows us to constrain the properties of the cloud via two routes. Firstly, we have the pressure at which the cloud's optical depth passes unity, i.e. the cloud top. Secondly, we have the retrieved exponent of the power-law that we have used for our simplified cloud opacity, which tells us roughly how ``non-grey" the cloud is. We have not strongly constrained the single scattering albedo (although higher single scattering albedo appears to be marginally preferred) so this does not help constrain the particle properties in this case.

Inspecting Figure~\ref{fig:2m2224thickprof} suggest thats both iron and corundum (Al$_{2}$O$_{3}$) condensates would be stable on our profile at the pressure of the cloud deck in 2M2224-0158, whilst Figure~\ref{fig:2m0500thickprof} suggests that only corundum would be stable at the 2M0500+0330 cloud deck. The silicates, enstatite and fosterite, on the other hand would not be stable at the cloud deck for either object. Our first conclusion, then, is that the dominant source of cloud opacity shaping the near-infrared spectra of these mid-L~dwarfs  is not silicate, but rather corundum and/or iron condensates.

This is a surprising result, since silicate condensates are widely regarded as the dominant source of cloud opacity in L~dwarf spectra. 
This is especially relevant for 2M2224-0158, for which a feature tentatively attributed to small silicate particles has been observed in the 8$\mu$m region \citep{cushing2006}. This feature was identified by a broad dip in the observed spectrum relative to the $T_{\rm eff} = 1700$K, $\log g = 5.0$ grid model (Figure~\ref{fig:longspec2}).  As can be seen, this model's spectrum does not match the shape of the observed spectrum for most of the SED, so it is difficult place the poor match at 8$\mu$m in context.  None-the-less, our results do not exclude the presence of silicate clouds. Our thermal profiles cross the stability curve for fosterite, so it is entirely possible that such clouds exist at lower pressures than the principal cloud deck, and are responsible for the feature seen at 8$\mu$m. 
As we saw in Section~\ref{sec:fakedata}, our retrieval is not sensitive to the vertical distribution of cloud above the level at which it becomes optically thick. So, while silicates may contribute to the cloud opacity, the dominant contribution is more likely attributable to a very optically thick cloud deck at location most consistent with corundum and/or iron rich condensates.  Future retrieval studies incorporating longer wavelength data (e.g. from the James-Webb Space Telescope) will resolve this question.

\begin{figure*}
\hspace{-0.8cm}
\includegraphics[width=350pt]{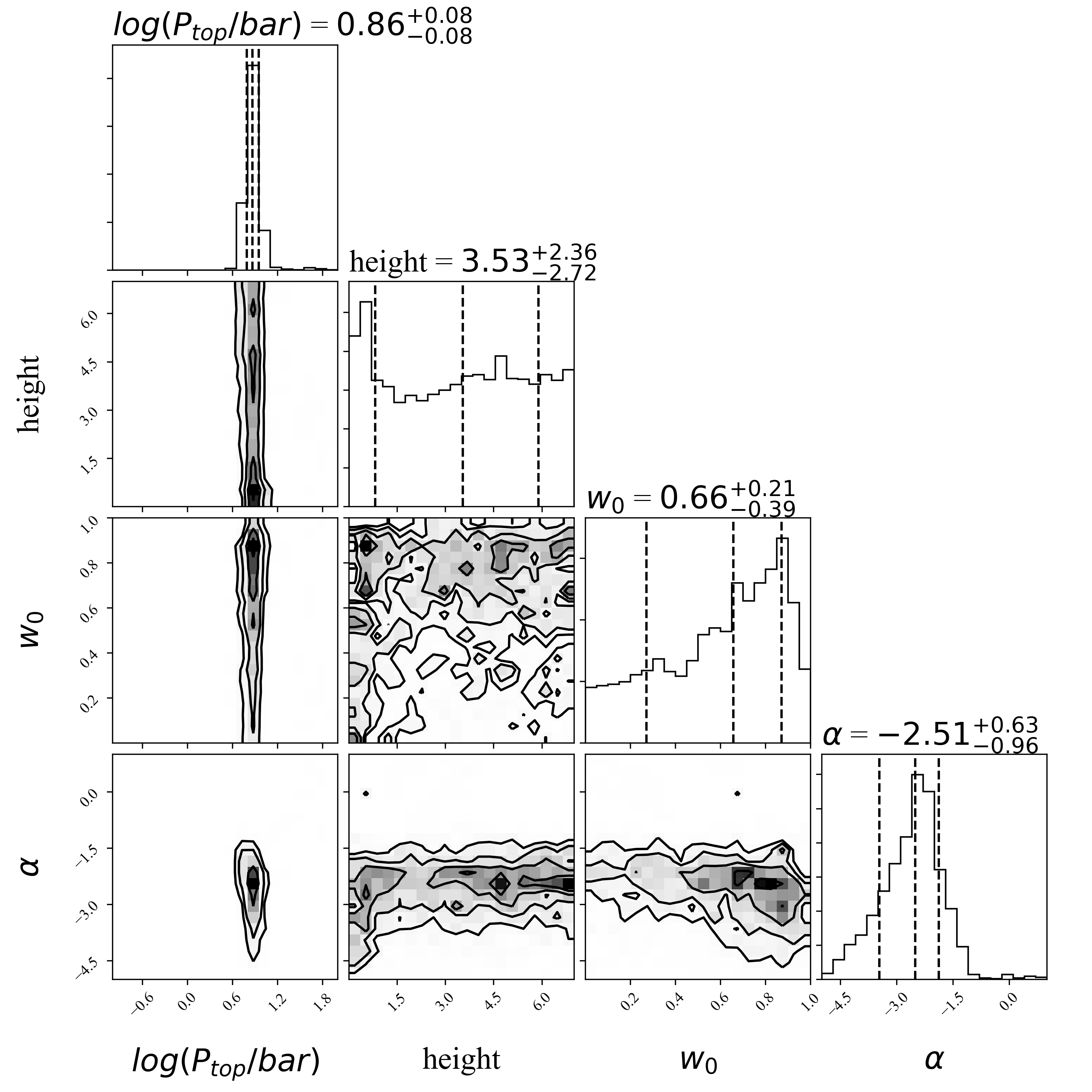}
\caption{Posterior probability distributions for some our retrieved cloud parameters in the selected cloud deck case for 2M0500+0330.
\label{fig:2m0500cloud}}
\end{figure*}

\begin{figure*}
\hspace{-0.8cm}
\includegraphics[width=350pt]{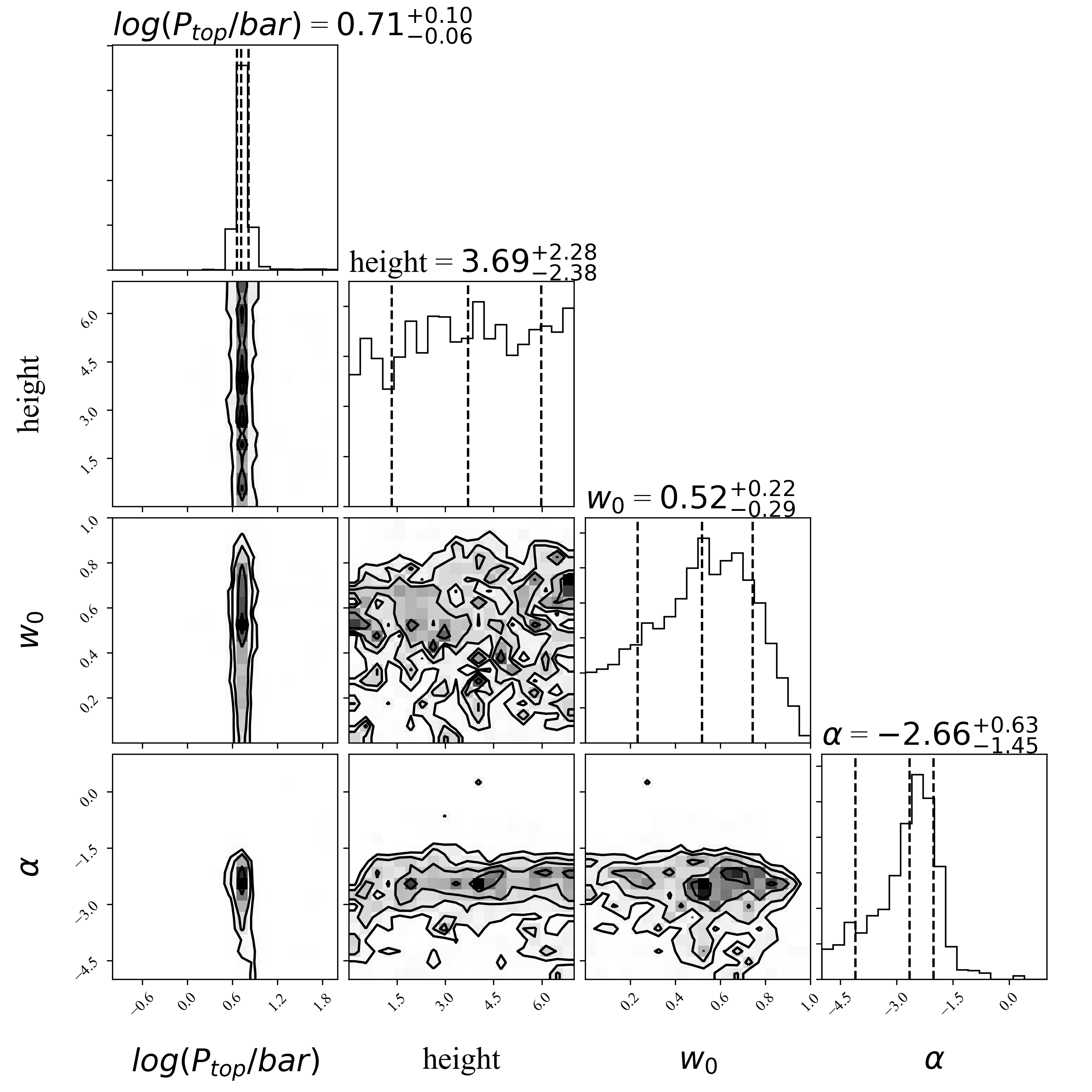}
\caption{Posterior probability distributions for some of our retrieved cloud parameters in the selected cloud deck case for 2M2224-0158. 
\label{fig:2m2224cloud}}
\end{figure*}

Our simple parameterisation of the cloud opacity precludes the detection of species-specific absorption features due to Mie scattering. However, the exponent ($\alpha$) of the power-law opacity that we estimated ($\tau \propto \lambda^\alpha$, $\alpha = -2.45^{+0.63}_{-0.96}$ and $\alpha = -2.7^{+0.63}_{-1.45}$ ; see Figures~\ref{fig:2m0500cloud} and~\ref{fig:2m2224cloud}) corresponds to a relatively steep reddening law, consistent with a cloud dominated by small, sub-micron, sized particles. We have attempted to reconstruct this opacity curve using Mie coefficients calculated for corundum and iron assuming a log-normal size distribution.  Even for a very narrow log-normal distribution centred on sub-micron particles, the impact of large grains in the distribution tail tend to grey-out the opacity and we are unable to reproduce the red opacity seen in our retrieval. A Hansen distribution \citep{hansen1971}, however, is able to reproduce our retrieved cloud opacity curve for a mean effective radius $\sim 0.1$~$\mu$m, and an effective variances $\sim 1$ for both iron and corundum.  The Hansen distribution was also preferred by \citet{hiranaka2016} for reddening the spectra of field L~dwarfs to match red L~dwarfs, and is commonly used to describe particle distributions in terrestrial water clouds.  It is worth noting that the forward grid models of \citet{sm08}, assume a log-normal particle size distribution for their clouds, and find particle sizes of order 10$\mu$m.

\section{Discussion and Conclusions}
\label{sec:conc}

Our first application of the retrieval technique to the cloudy atmospheres of two mid-L~dwarfs has revealed a number of interesting features. Most significantly, both of our targets appear to have thermal profiles that are several hundred degrees cooler at depth than predicted by forward grid models for similar $T_{\rm eff}$. In addition, both targets display warmer upper atmospheres than predicted by the same models.  Our retrieved temperature in the upper atmosphere of 2M2224-0158 is in close agreement with that estimated by \citet{sorahana2014}. Our retrieval of a similar profile for the L4 spectral template 2M0500+0330, suggests that the mechanism responsible for heating the upper atmosphere of mid-L~dwarfs may be widespread. 

We confirmed that models which include clouds are be preferred over cloud-free models. Moreover, an optically thick, non-grey, cloud deck is preferred over models with either grey or slab clouds. We have constrained the location of the cloud deck to near the 5 bar level in both cases. The temperature at this pressure is too high for stable silicate clouds and we conclude that the dominant cloud opacity arises from either iron condensate, corundum, or some combination of the two. In addition, we find that the non-grey opacity is suggestive of a cloud dominated by small particles. 

 Our cloud-free retrieval results are shown in the Appendix. Their profiles display similar morphology to that seen in Test Case 2 where an isothermal profile is substituted for the presence of a cloud. This bears resemblance to the suggestion by \citet{tremblin2015,tremblin2016} that a sufficiently shallow thermal profile, driven by non-equilibrium chemistry, can reproduce the spectral shape of L~dwarfs.

\begin{figure}
\hspace{-0.8cm}
\includegraphics[width=290pt]{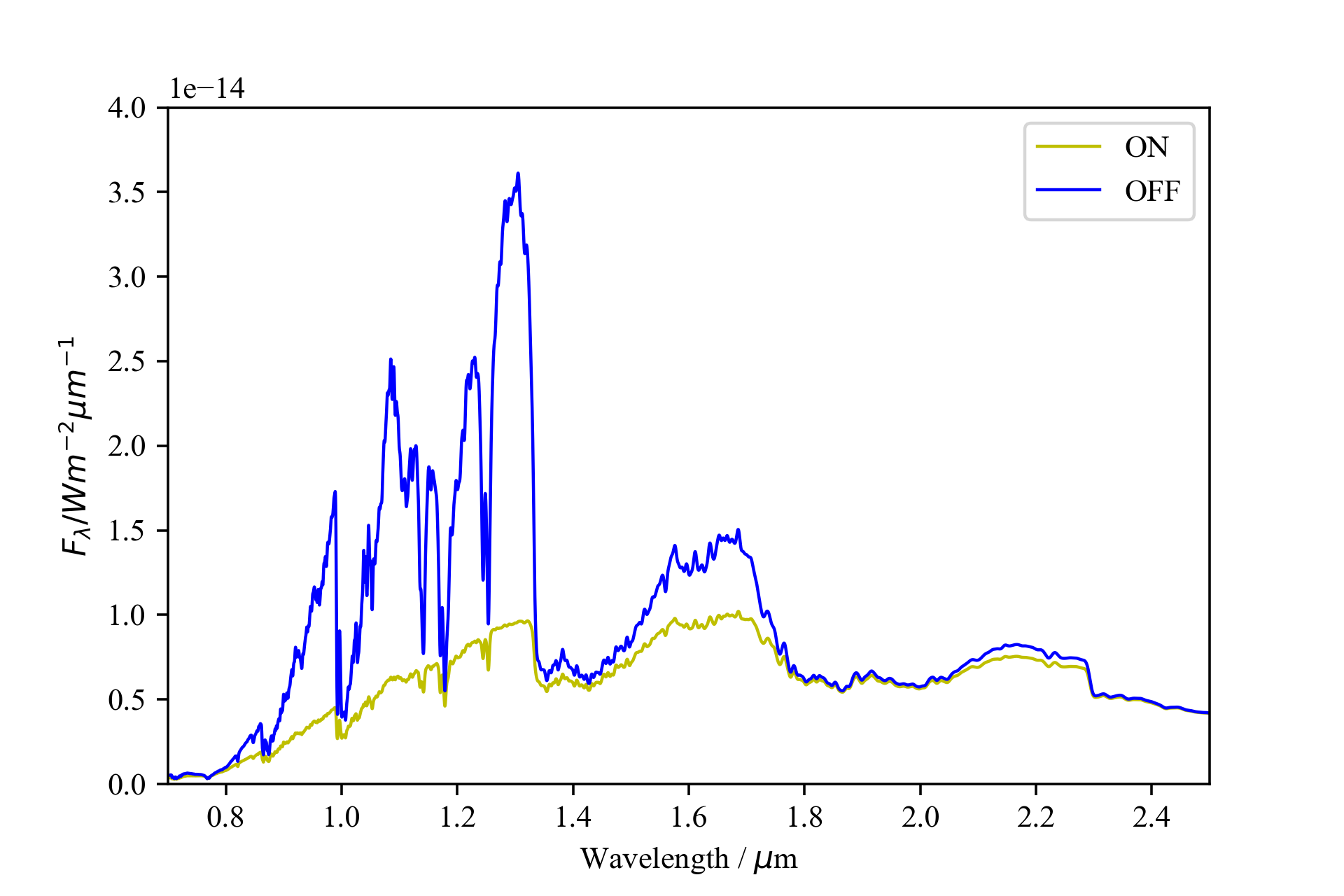}
\caption{A comparison demonstrating the impact of the cloud deck opacity for our maximum-likelihood retrieval for the 2M2224-0158. All parameter are identical between the spectra, but one has the clouds turned on, and one turned off. 
\label{fig:clouddemo}}
\end{figure}

It is interesting to note that, even with the inclusion of an optically thick, non-grey, cloud-deck, our cloudy retrieved profiles for both 2M2224-0158 and 2M0500+0330 have shallower gradients than the predictions of the grid models at lower pressures than $\sim 1$~bar.  As noted earlier, this would require some form of stratospheric heating mechanism. 
The confidence range for the deeper profiles are consistent with them being essentially parallel to the grid model profiles (i.e. following the same adiabat). However, the profile for 2M2224-0158 appears marginally shallower, and both sets of confidence intervals allow even shallower gradients. We can thus not rule out that a mechanism such as that proposed by \citet{tremblin2015,tremblin2016} might be at work alongside the influence of thick clouds to shape the SEDs of these L~dwarfs. If present, however, these effects appear to be secondary  to the dramatic impact of the clouds, as illustrated in Figure~\ref{fig:clouddemo}.

We also note that our ability to accurately constrain $T_{\rm eff}$ using just the near-infared spectra of these targets is impressive, and the close agreement between the extended wavelength predictions based on our retrieval and WISE photometry for these targets suggests our retrieved properties are, on the whole, sound.

Our retrieved gas abundances are broadly consistent with the expectations from our thermochemical model with a couple of exceptions. The striking disagreement between our retrieved CO abundance and that predicted by our thermochemical model for solar abundances, is concerning. This, coupled with the poor agreement between grid model spectra and the observed CO band head in the $K$ band suggests there may be a widespread problem with fitting this absorption feature which may reflect a hitherto unidentified problem with widely used CO opacities. Given its importance for deriving C/O ratios for directly imaged exoplanets \citep[e.g.][]{konopacky2013}, this is a key problem to solve over the coming years. More rigorous validation and testing of our ability to retrieve accurate gas abundances will be possible in future work where we will focus on benchmark L~dwarfs for which detailed compositional constraints can be obtained via association to a primary star.

\appendix
\section{G570D results}
\label{sec:g570}
\begin{figure}
\hspace{-0.8cm}\includegraphics[width=290pt]{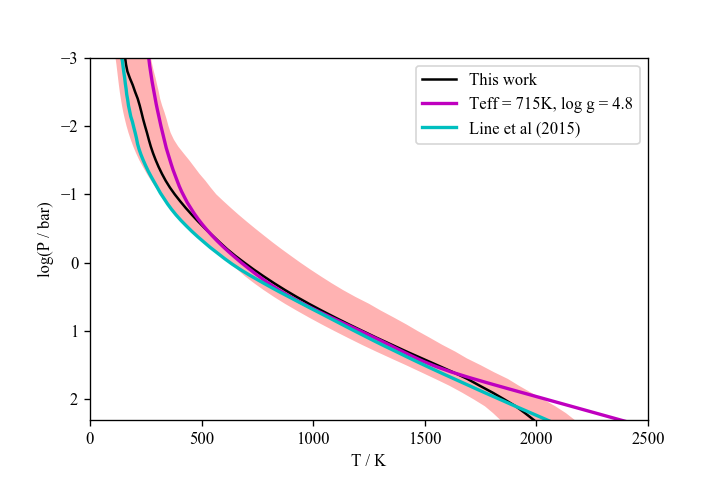}
\caption{Our retrieved thermal profile for G570D. The median case thermal profile is shown as a solid black line, with the 1$\sigma$ intervals is indicated with red.  Also plotted are the retrieved profile and corresponding grid profile as shown in \citet{line2015}.
\label{fig:g570TP}}
\end{figure}

\begin{figure*}
\hspace{-0.8cm}
\includegraphics[width=550pt]{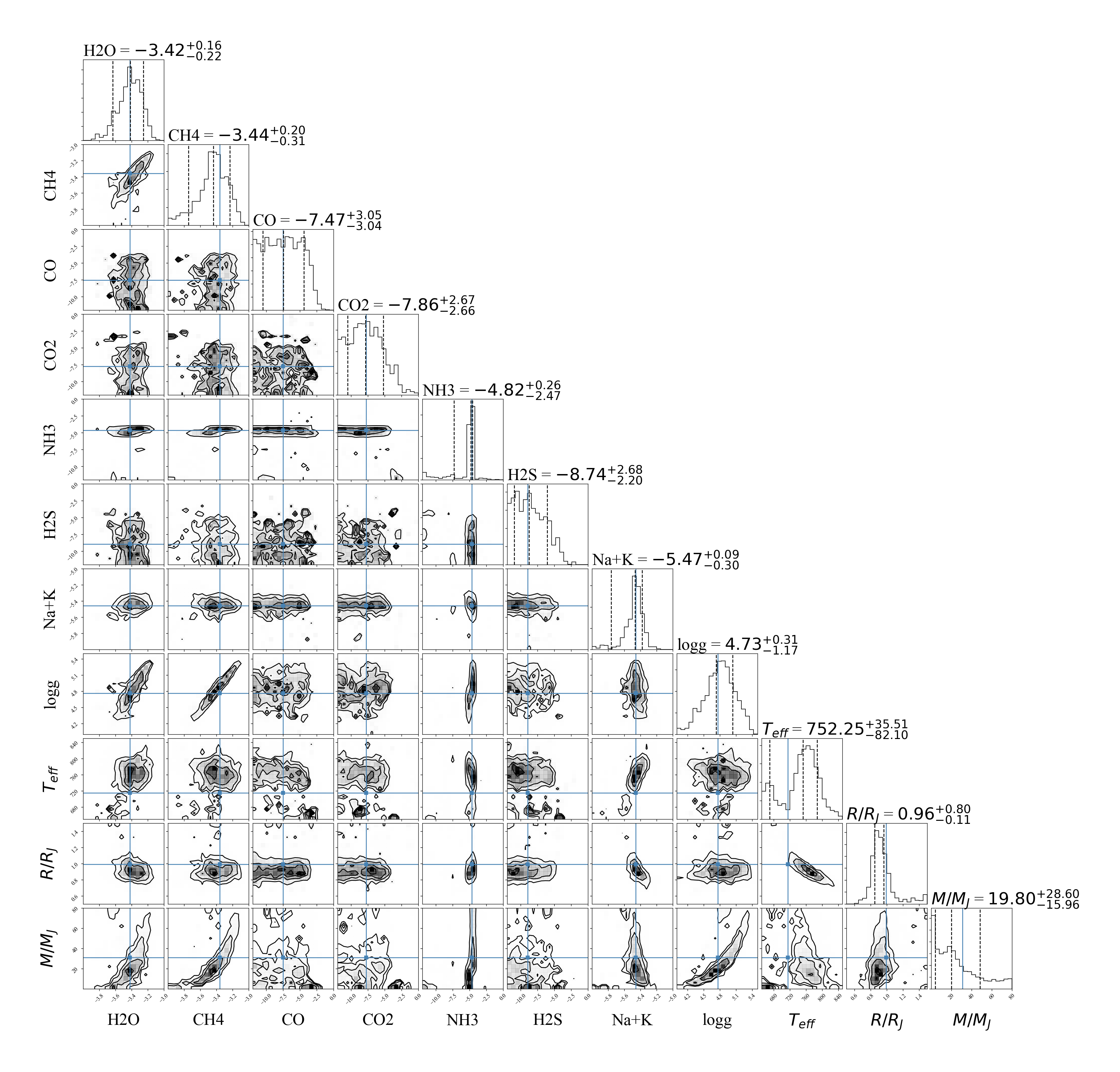}
\caption{Posterior probability distributions for some our retrieved parameters G570D. Also shown are $M$, $R$ and $T_{\rm eff}$ which are not directly retrieved, but are inferred from our retrieved parameters $R^2/D^2$, $\log g$ and predicted spectra. The values found by \citet{line2015} are shown with blue dashed lines. 
\label{fig:g570post}}
\end{figure*}

Here we present our posterior distributions for the retrieval of the properties of G570D. In Figure~\ref{fig:g570TP} we show the retrieved thermal profile using the 5-parameter approximation of \citet{madhu2009} compared to the retrieved profile from \citet{line2015} and their corresponding grid model. In Figure~\ref{fig:g570post} we show our posterior distributions for the rest of the parameters, again compared the retrieved values from \citet{line2015}.

\section{Contribution Functions}
\label{sec:cont}
\begin{figure}
\hspace{-0.8cm}\includegraphics[width=290pt]{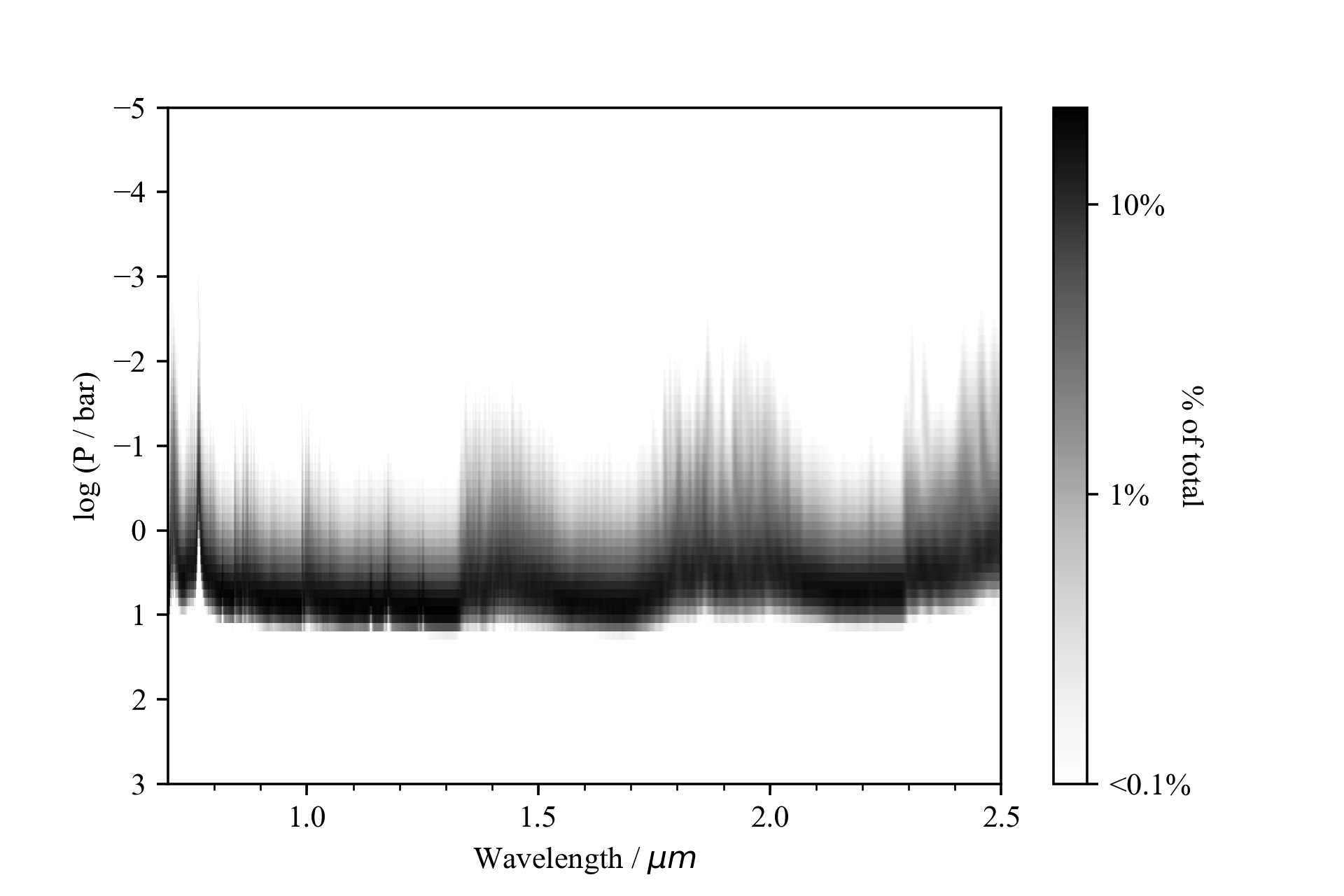}
\caption{ The contribution function for the maximum likelihood retrieval of 2M0500+0330 under the non-grey, optically thick cloud deck assumption. The grey-scale indicates the relative contributions of 0.1 dex thick layers.  
\label{fig:2m0500cont}}
\end{figure}

\begin{figure}
\hspace{-0.8cm}\includegraphics[width=290pt]{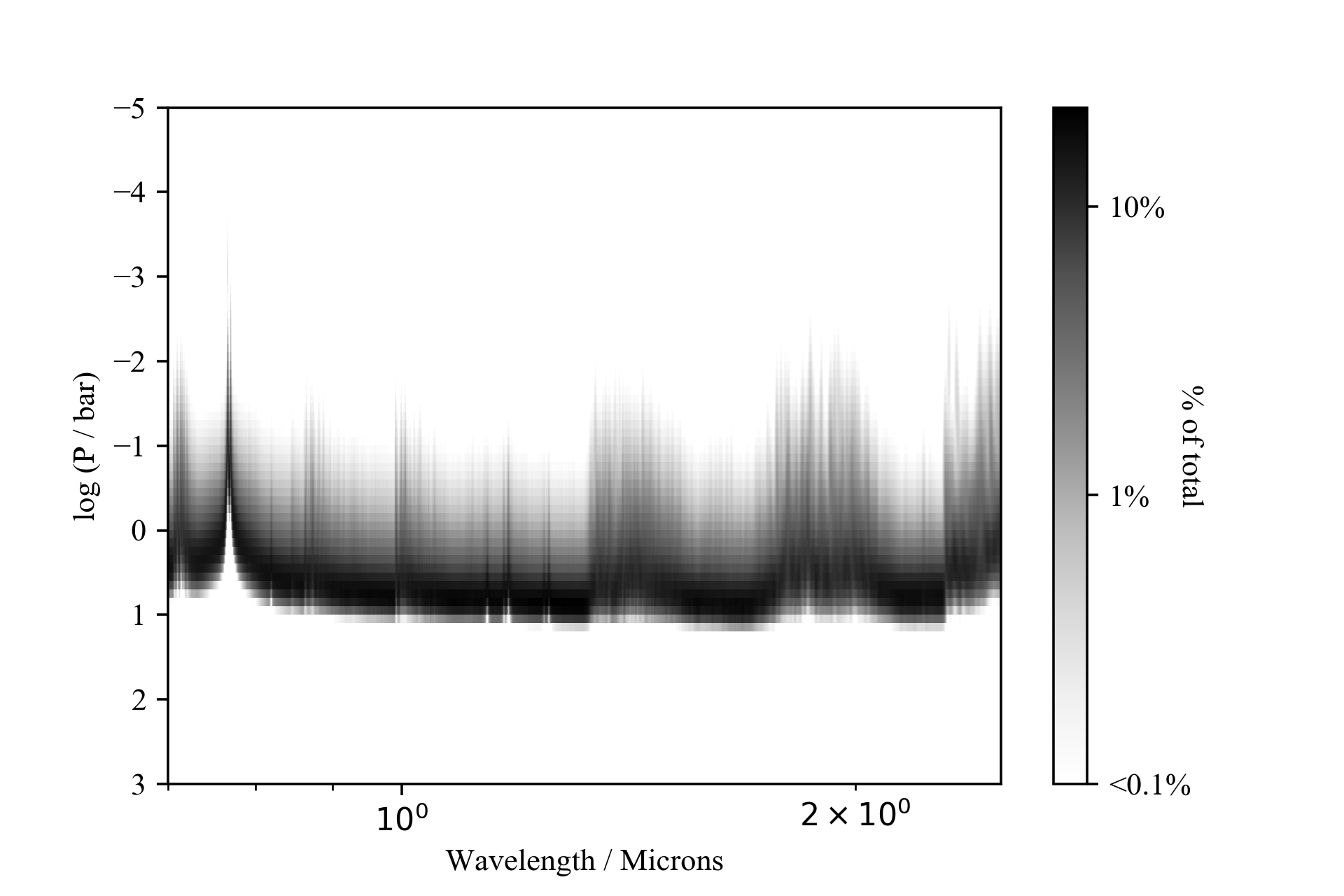}
\caption{ The contribution function for the maximum likelihood retrieval of 2M2224-0158 under the non-grey, optically thick cloud deck assumption. The grey-scale indicates the relative contributions of 0.1 dex thick layers.  
\label{fig:2m2224cont}}
\end{figure}

The contribution functions for the retrievals under the non-grey optically thick cloud deck assumption are shown in Figures~\ref{fig:2m0500cont} and~\ref{fig:2m2224cont}.

\section{Cloud-free retrievals results}
\label{sec:NC}

\begin{figure}
\hspace{-0.8cm}\includegraphics[width=290pt]{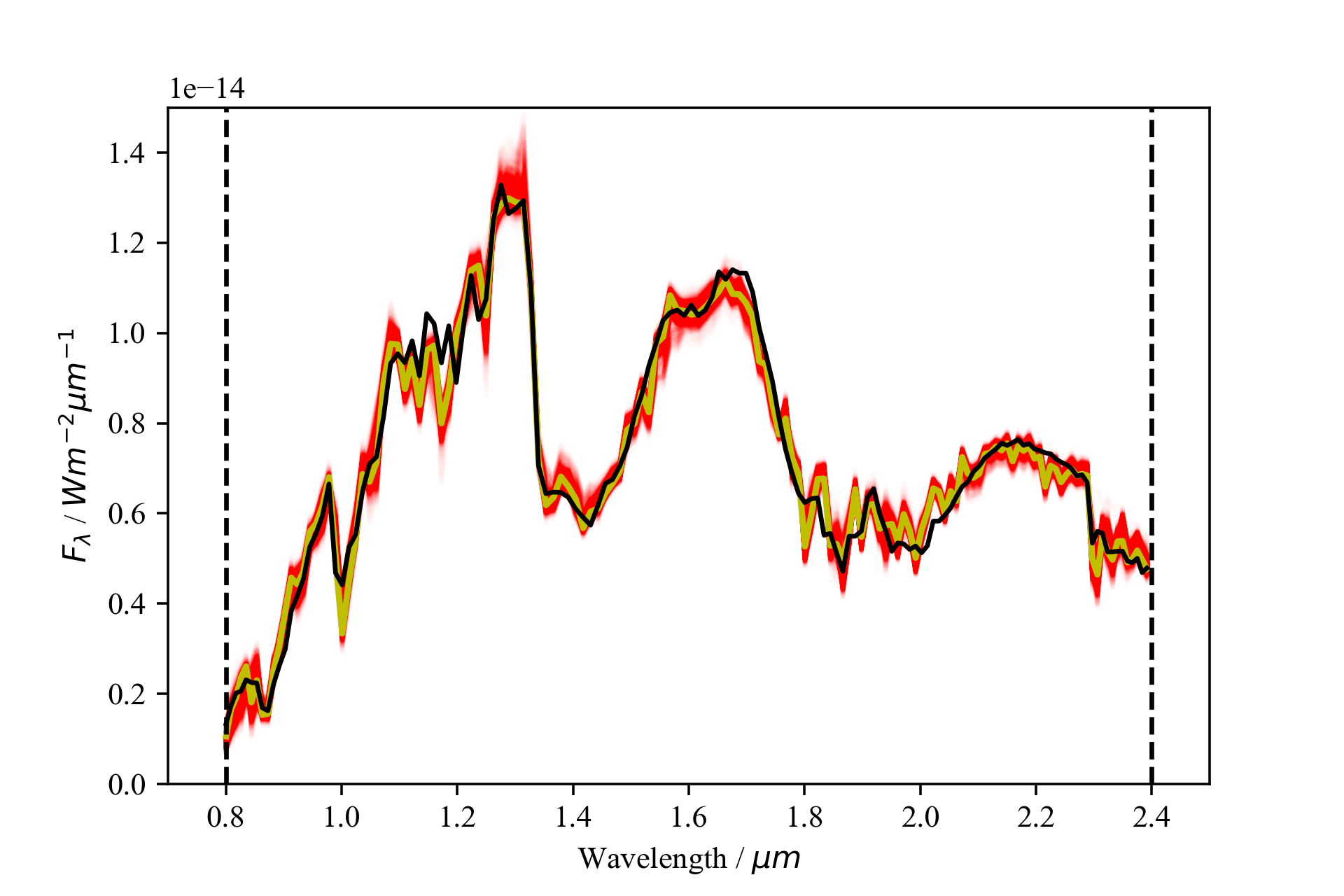}
\caption{ The 2M0500+0330 forward model output spectra for 5000 random draws from our final 2000 iterations (comprising 544000 independent samples) of the {\sc emcee} sampler for the cloud-free case , shown in red. The maximum likelihood case is plotted in yellow, and the Spex spectrum for our target is plotted in black \citep{gagliuffi2014}. 
\label{fig:2m0500NCspec}}
\end{figure}

\begin{figure}
\hspace{-0.8cm}\includegraphics[width=290pt]{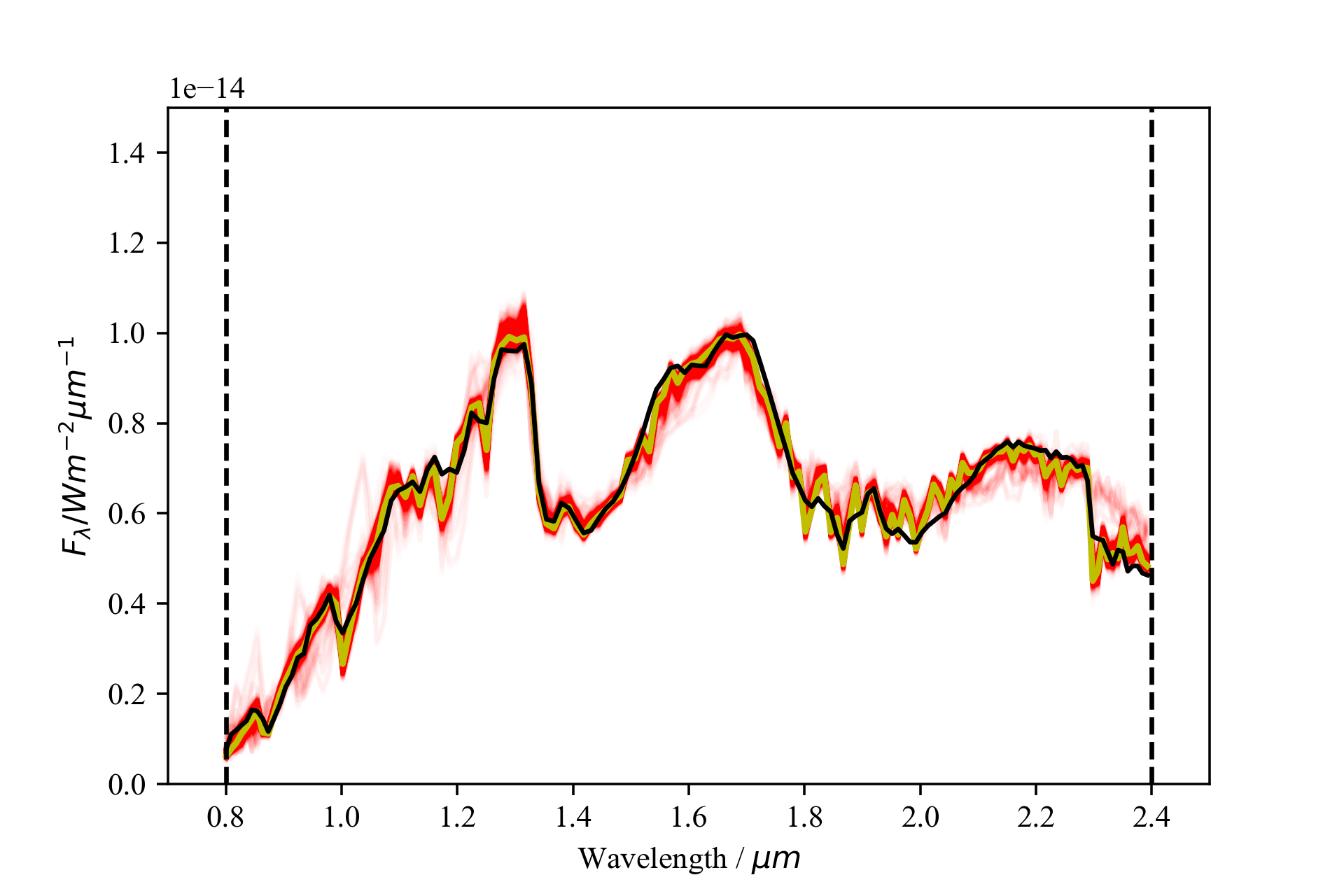}
\caption{ The 2M2224+0158 forward model output spectra for 5000 random draws from our final 2000 iterations (comprising 544000 independent samples) of the {\sc emcee} sampler for the cloud-free case , shown in red. The maximum likelihood case is plotted in yellow, and the Spex spectrum for our target is plotted in black \citep{burgasser2010a}. 
\label{fig:2m2224NCspec}}
\end{figure}

\begin{figure}
\hspace{-0.8cm}\includegraphics[width=290pt]{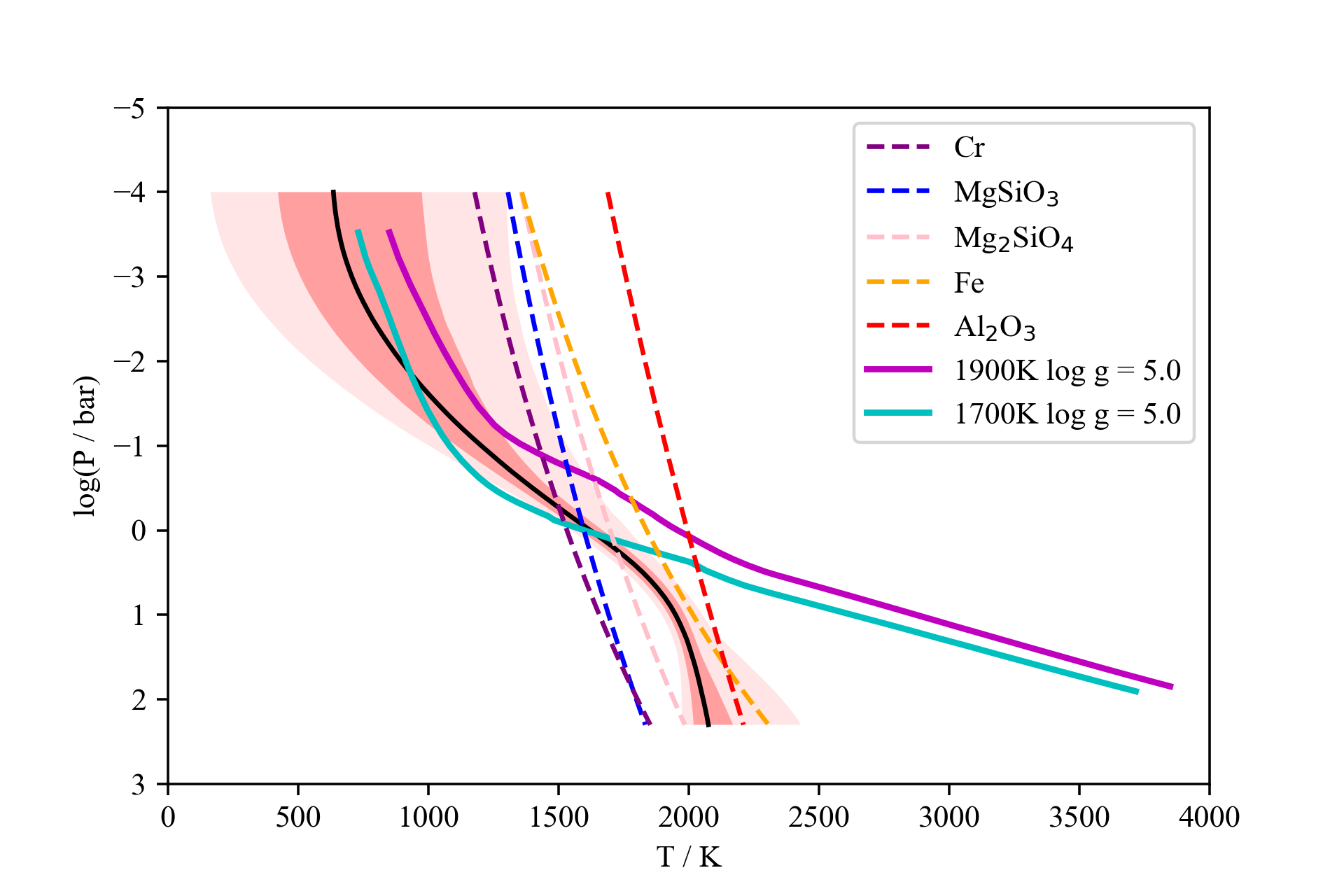}
\caption{Our retrieved thermal profile for the cloud-free model for 2M0500-0330. The median case thermal profile is shown as a solid black line, with the 1$\sigma$  and 2$\sigma$ intervals are indicated with red and pink shading.  Also plotted are condensation curves for likely cloud species, and the forward model grid (T,P) profiles for $f_{sed} = 3$, $\log g = 5.0$,  for $T{\rm eff} =  1700$K and 1900~K, which bracket the $T_{\rm eff}$ estimated by \citet{filippazzo2015}, and that inferred from this retrieval model.
\label{fig:2m0500NCprof}}
\end{figure}

\begin{figure}
\hspace{-0.8cm}\includegraphics[width=290pt]{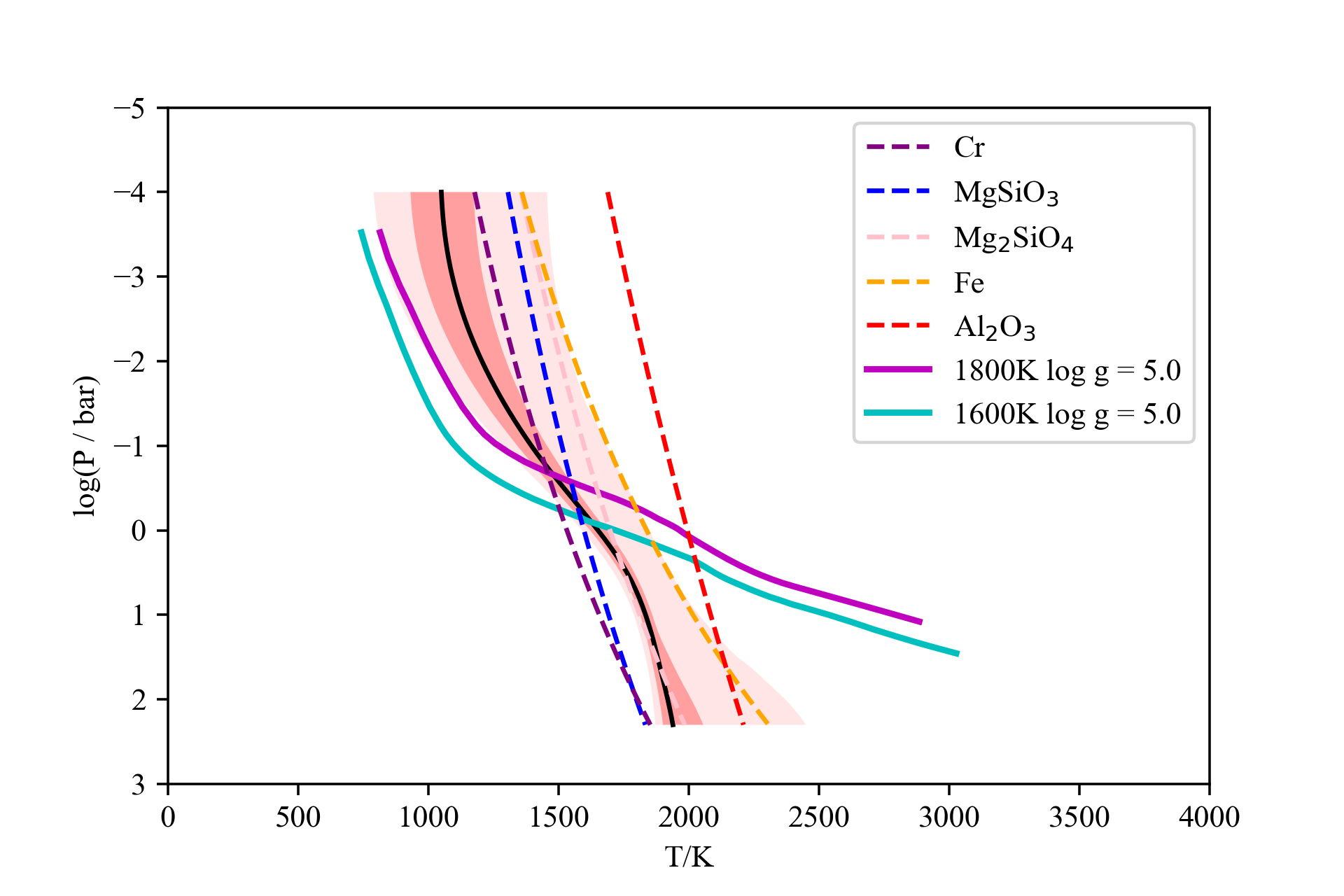}
\caption{Our retrieved thermal profile for the cloud-free model for 2M2224-0158. The median case thermal profile is shown as a solid black line, with the 1$\sigma$  and 2$\sigma$ intervals are indicated with red and pink shading.  Also plotted are condensation curves for likely cloud species, and the forward model grid (T,P) profiles for $f_{sed} = 3$, $\log g = 5.0$,  for $T{\rm eff} =  1600$K and 1700~K, which bracket the $T_{\rm eff}$ estimated by \citet{filippazzo2015}, and that inferred from this retrieval model.
\label{fig:2m2224NCprof}}
\end{figure}

\begin{figure*}
\hspace{-0.8cm}
\includegraphics[width=550pt]{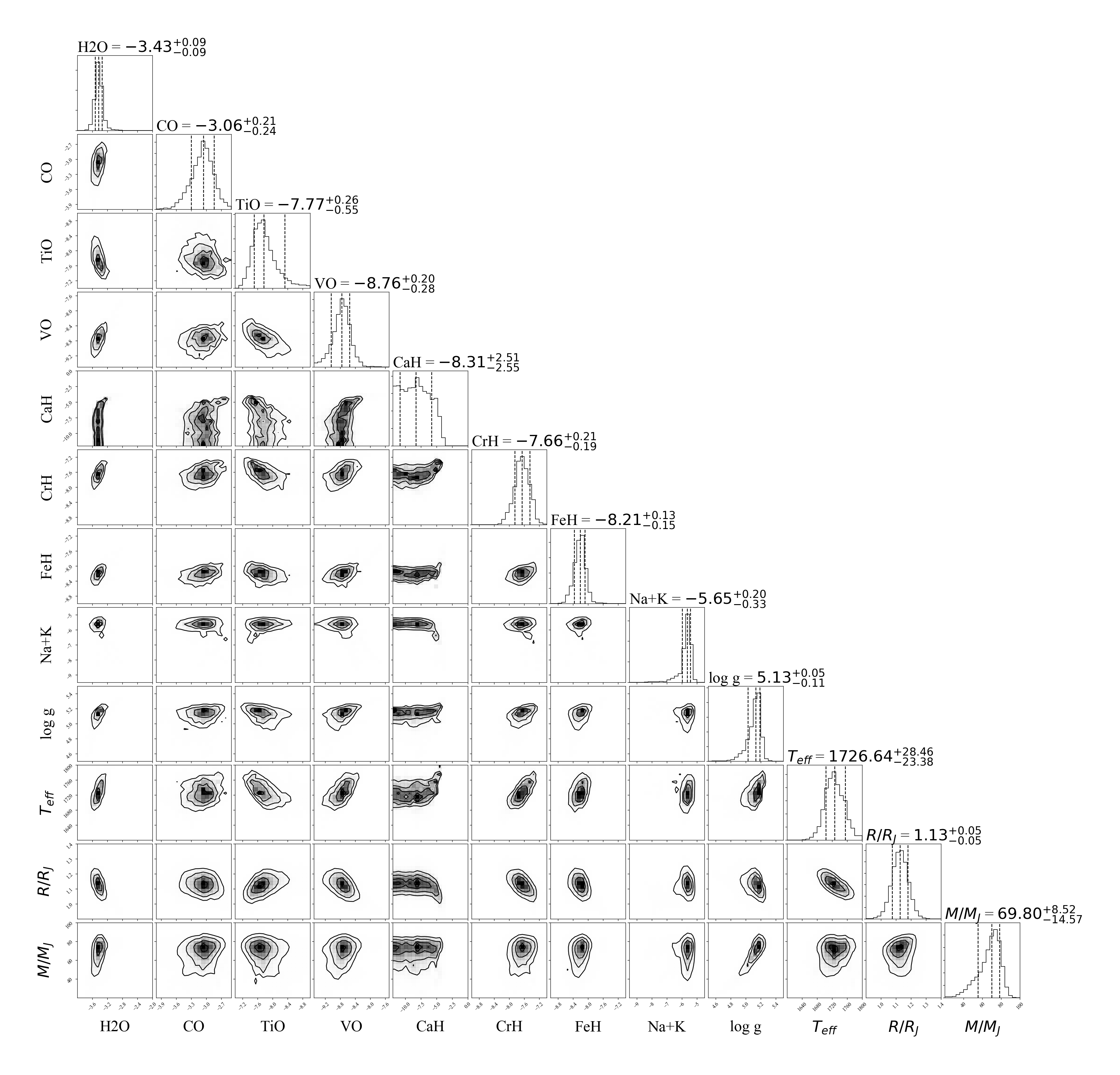}
\caption{Posterior probability distributions for some our retrieved parameters in the cloud-free case for 2M0500+0330. Also shown are $M$, $R$ and $T_{\rm eff}$ which are not directly retrieved, but are inferred from our retrieved parameters $R^2/D^2$, $\log g$ and predicted spectra. 
\label{fig:2m0500post_NC}}
\end{figure*}

\begin{figure*}
\hspace{-0.8cm}
\includegraphics[width=550pt]{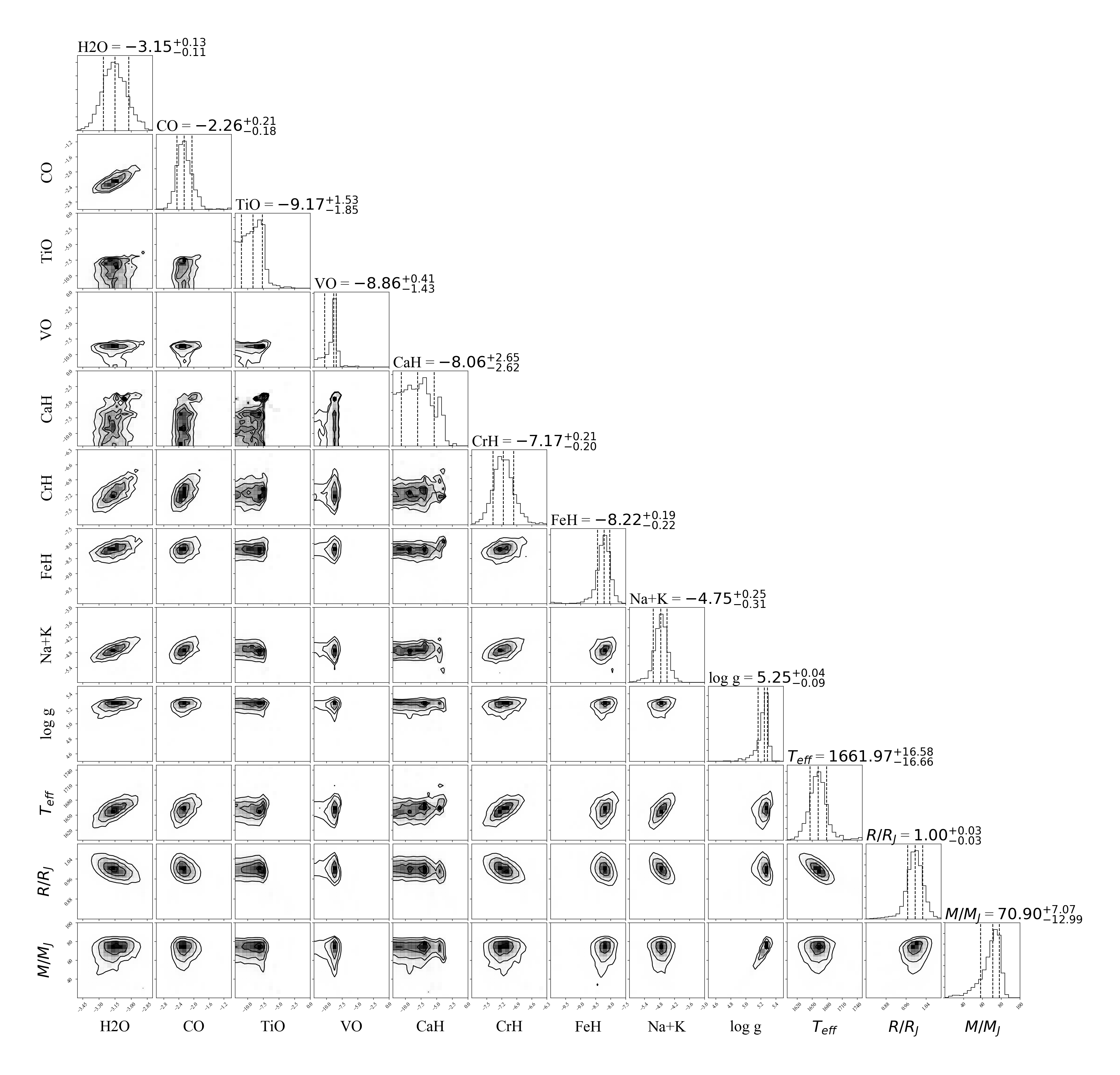}
\caption{Posterior probability distributions for some of our retrieved parameters in the cloud-free case for 2M2224-0158. Also shown are $M$, $R$ and $T_{\rm eff}$ which are not directly retrieved, but are inferred from our retrieved parameters $R^2/D^2$, $\log g$ and predicted spectra. 
\label{fig:2m2224post_NC}}
\end{figure*}

Figures~\ref{fig:2m0500NCspec} and~\ref{fig:2m2224NCspec} show the random draws from the posterior and maximum likelihood spectra for the cloud-free retrieval cases for 2M0500+0330 and 2M2224-0158 respectively. Figures~\ref{fig:2m0500NCprof} and~\ref{fig:2m2224NCprof} show their retrieved thermal profiles, while Figures~\ref{fig:2m0500post_NC} and~\ref{fig:2m2224post_NC} show the posterior distributions for the retrieved parameters and derived properties under the same assumption. 

\begin{figure*}
\hspace{-0.8cm}
\includegraphics[width=350pt]{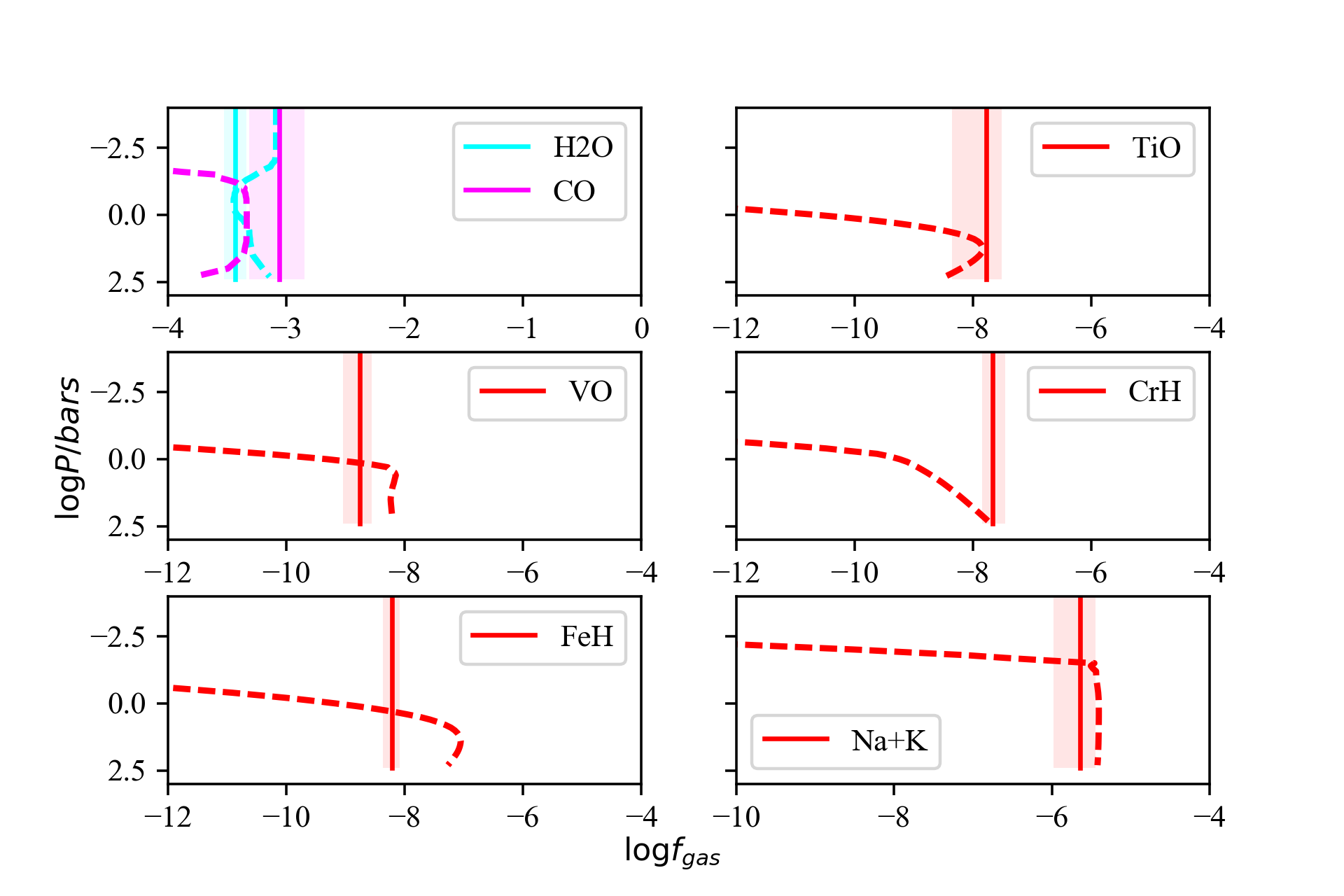}
\caption{Retrieved abundances for 2M0500+0330 and predicted thermochemical equilibrium abundances for our retrieved profile under the cloud-free assumption. 
\label{fig:2m0500abund_nc}}
\end{figure*}

\begin{figure*}
\hspace{-0.8cm}
\includegraphics[width=350pt]{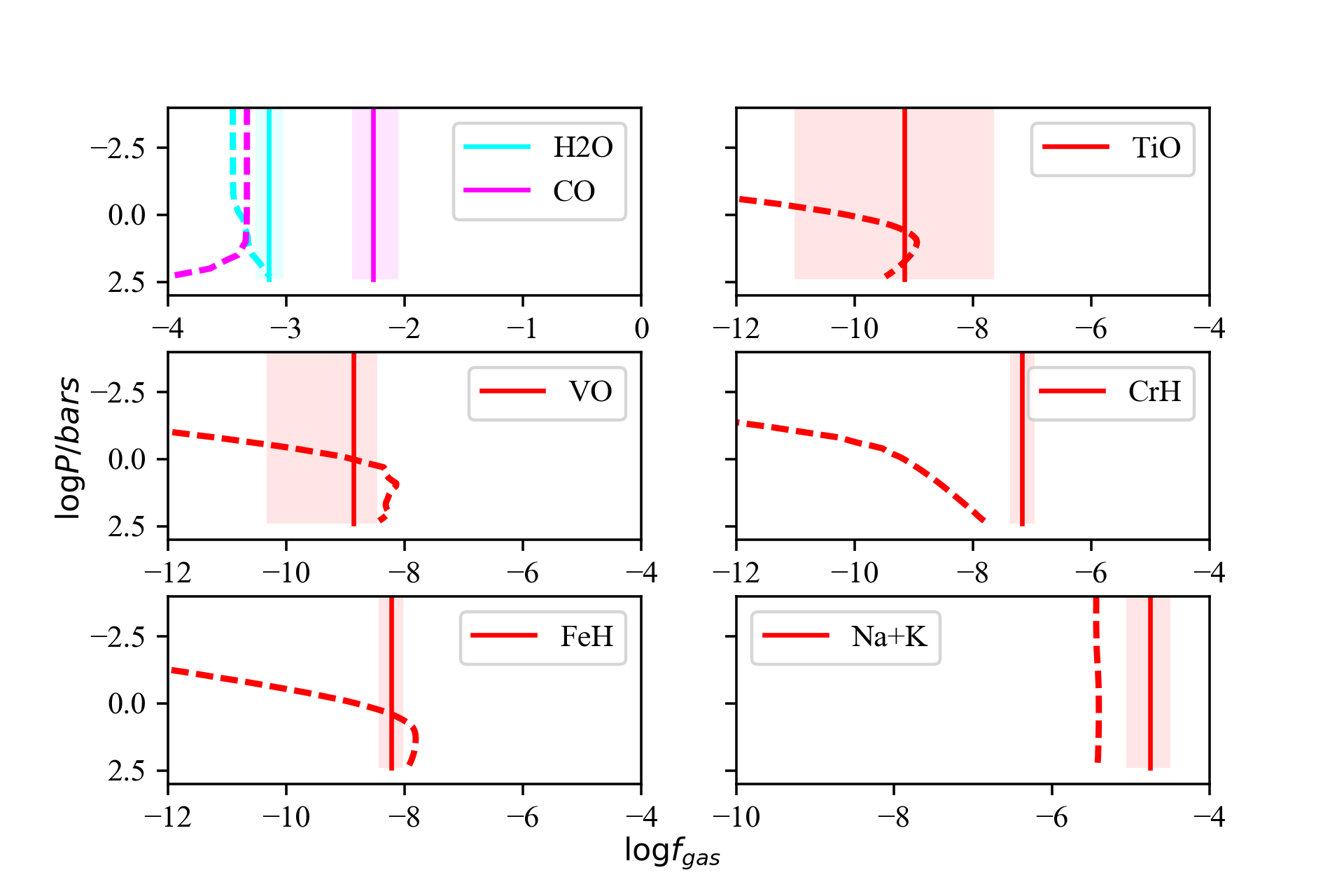}
\caption{Retrieved abundances for 2M2224-0158 and predicted thermochemical equilibrium abundances for our retrieved profile under the cloud-free assumption. 
\label{fig:2m2224abund_nc}}
\end{figure*}

\section*{Acknowledgements}
Thanks to Stuart Littlefair for helpful discussions during the preparation of this manuscript. BB acknowledges financial support from the European Commission in the form of a Marie Curie International Outgoing Fellowship (PIOF-GA-2013- 629435). CV acknowledges support by NSF grant AST-1312305. DS is supported in part by a NASA Astrophysics Theory Program grant.
 
\bibliographystyle{mn2e}
\bibliography{refs}

\end{document}